\documentclass[twocolumn,prl,superscriptaddress]{revtex4-2}
\usepackage[utf8]{inputenc} 
\usepackage{amsmath,amssymb,mathrsfs}
\usepackage{physics}
\usepackage{tikz}
\usepackage{tikz-feynman}

\tikzset{
	cross/.style={path picture={
			\draw[black]
			(path picture bounding box.south east) -- (path picture bounding box.north west)
			(path picture bounding box.south west) -- (path picture bounding box.north east);
	}}
}

\usepackage[hidelinks]{hyperref}

\usepackage{transparent}

\definecolor{gbcolor}{rgb}{.8,.3,.1}

\newcommand{\eq}[1]{(\ref{#1})}

\begin{document}
%
%

\title{Classical and quantum evolution of inflationary fluctuations}

\author{Guillermo Ballesteros}
\email{guillermo.ballesteros@uam.es}
\affiliation{Departamento de F\'{\i}sica Te\'{o}rica, Universidad Aut\'{o}noma de Madrid (UAM), Campus de Cantoblanco, 28049 Madrid, Spain \vspace{0.06cm}}
\affiliation{Instituto de F\'{\i}sica Te\'{o}rica UAM-CSIC,  Campus de Cantoblanco, 28049 Madrid, Spain \vspace{0.06cm}}

\author{Jes\'us Gamb\'in Egea}
\email{j.gambin@csic.es}
\affiliation{Instituto de F\'{\i}sica Te\'{o}rica UAM-CSIC,  Campus de Cantoblanco, 28049 Madrid, Spain \vspace{0.06cm}}

\author{Alejandro P\'erez Rodr\'iguez  \vspace{0.15cm}} 
\email{aperezro@uw.edu}
\affiliation{Department of Physics, University of Washington, Seattle, WA 98195, USA}

\begin{abstract}

We compare the correlation functions of inflationary perturbations computed either with quantum or classical dynamics. Even if they are enforced to agree at
a specific time during inflation, classical and quantum correlations will differ at the end of inflation, provided that interactions are relevant. The difference between the results of the classical and quantum computations is exponentially sensitive to the number of e-folds elapsed from the time of agreement. We illustrate this finding with the tree-level bispectrum of the primordial curvature fluctuation and the one-loop power spectrum of tensor modes. We also show that classical evolution from a finite time does not imply the appearance of poles in the scalar bispectrum. 

\end{abstract}

\maketitle

\section{Introduction}

Proposed in the eighties, inflation \cite{Starobinsky:1979ty,Guth:1980zm,Linde:1981mu} is the most compelling explanation for the flatness and homogeneity of the Universe at cosmological distances, as well as for the properties of the primordial fluctuations inferred from the Cosmic Microwave Background (CMB) \cite{WMAP:2003syu,Senatore:2009gt,Planck:2018jri,Planck:2019kim} and the large scale structure (LSS) distribution of galaxies \cite{Cabass:2022wjy,DAmico:2022gki}. Yet, inflation is still unproven. Detecting primordial gravitational waves from their effect on the polarization of the CMB B-modes is often regarded as the would-be definitive proof that inflation did actually happen \cite{Starobinsky:1979ty} or, as it is sometimes put, of the quantum nature of gravity~\cite{Achucarro:2022qrl}. However, if inflation occurred at a low enough energy scale, this path may always elude us. Future experiments \cite{LiteBIRD:2022cnt} aim to measure the tensor-to-scalar ratio at the level of $10^{-3}$, corresponding to a Hubble scale of nearly $10^{13}$~GeV. Lowering the observational sensitivity to the Hubble
scale by just an order of magnitude represents a tremendous futuristic challenge. For this reason, it is important to ponder whether other properties of inflation may offer different handles to test it.

{A possibly fruitful avenue may open up by exploring its implications at cosmologically small distances that cannot be probed by CMB and LSS data. 
While inflation is required to last around fifty or sixty e-folds to solve the flatness and horizon problems it was postulated for \cite{Liddle:2003as}, CMB and LSS data can probe roughly only the first decade of them. The properties of the primordial spectra at small cosmological scales are unknown and only crude upper limits on the scalar amplitude exist from the non-observation of primordial black holes (PBH), see e.g.~\cite{Gow:2020bzo}. 
The latter have become a central target in the last decade because they may constitute the dark matter and probe scales inaccessible with CMB and LSS. 
If inflation led to abundant PBH production, it should also have generated a stochastic background of second-order scalar-induced gravitational waves \cite{Saito:2008jc,Bartolo:2018evs}. Both this background and PBH themselves therefore provide complementary probes of small-scale inflation. However, a cosmologically relevant PBH abundance requires a scalar spectrum far larger than at CMB scales, potentially pushing the standard perturbative treatment of inflationary {quantum} fluctuations beyond its regime of validity.

This has motivated the use of {\it classical} lattice simulations to study inflation in mildly non-linear and non-perturbative regimes
\cite{Caravano:2021pgc,Caravano:2024tlp,Caravano:2024moy,Caravano:2026hca}. 
In this approach, the inflaton and any additional fields are discretized and evolved according to their classical equations of motion. At the moment, no other numerical method exists in the purely non-perturbative regime. In this Letter, we challenge this approach by asking whether these simulations are capable of reproducing the true quantum nature of inflation. In order to do so, we compare the outcomes of classical and quantum perturbation theories in the perturbative regimes, finding a broadly general discrepancy. 

{We show that even if stochastic classical initial conditions are chosen to match the quantum statistics of the inflationary fluctuations at a finite time, their subsequent classical evolution generically yields primordial correlators that differ from the standard quantum predictions.}
{This obstructs the use of non-perturbative classical lattice evolution to compute inflationary correlators, as well as perturbative calculations based on classical Green functions.}
The classical evolution also retains memory of the matching time, which in quasi-de Sitter translates into a mismatch that grows exponentially with the number of e-folds elapsed before horizon crossing. We illustrate these points with the calculation of the primordial bispectrum of curvature fluctuations in a concrete model with a derivative interaction and with the inflationary scalar-induced tensor spectrum in de Sitter.

{Inflation is conventionally viewed as a framework that provides a quantum theory for primordial fluctuations. Their statistical properties, encoded in their correlation functions, are computed using quantum field theory methods
\cite{Maldacena:2002vr,Weinberg:2005vy,Senatore:2009cf,Chen:2017ryl,Arkani-Hamed:2018kmz} and, after inflation ends, can be regarded as the initial conditions for the Universe \cite{Baumann:2022jpr}.} 
Yet, it remains unclear how observations could identify the quantum nature of these fluctuations, see \cite{Kiefer:1998qe,Maldacena:2015bha,Campo:2005sv,Martin:2017zxs,Launay:2024trh,Ireland:2026txt} for several proposals. It was proposed in \cite{Green:2020whw} that poles in folded configurations of primordial correlators can be a diagnostic of classical, as opposed to standard (Bunch-Davies) quantum, fluctuations. As a by-product of our main analysis, we show that classical evolution from stochastic initial conditions imposed at any finite time does not produce such poles.
}

\section{Quantum and Classical Dynamics}

Let us consider a collection of canonically normalized real scalar fields, $\Phi_a$, $a=1, \ldots, N$, and their conjugate momenta, which in Fourier space satisfy $\Pi_{a,\vb{k}}(t)=\dot\Phi_{a,-\vb{k}}$. Let us also consider a quantum operator $\hat{\mathcal{O}}(t)$ and its classical analogue $\mathcal{O}(t)$, which depend, respectively, on quantum and classical versions of those fields.
Then, $\hat{\mathcal{O}}(t)$ and ${\mathcal{O}}(t)$ evolve in time as follows, see Appendix 1:
\begin{widetext}
	\begin{align}
			\hat{\mathcal{O}}\left(t \right) & = \hat{F}^\dagger(t,t_0) \hat{\mathcal{O}}_f(t) \hat{F}(t,t_0) = \sum_{n = 0}  \frac{1}{\left( i\hbar \right)  ^n} \int_{t_0}^t \dd t_1 \int_{t_0}^{t_1} \dd t_2 \cdots \int_{t_0}^{t_{n-1}} \dd t_n \,\comm{ \comm{\comm{\hat{\mathcal{O}}_f\left(t \right) }{ \hat{H}_I(t_1)}}{ \hat{H}_I(t_2)}}{ \cdots , \hat{H}_I(t_n)} \,, \label{eq: Dyn Q}\\
		\mathcal{O}\left(t \right) & = F(t,t_0) \mathcal{O}_f(t)= \sum_{n = 0} \int_{t_0}^t \dd t_1 \int_{t_0}^{t_1} \dd t_2 \cdots \int_{t_0}^{t_{n-1}} \dd t_n \, \{ \cdots\{\{ \mathcal{O}_f\left(t \right)  , H_I(t_1)\}_0, H_I(t_2)\}_0 \cdots , H_I(t_n)\}_0 \,. \label{eq: Dyn C}
	\end{align}
\end{widetext}
The operators $\hat{F}$ and $F$ are the time evolution generators in the interaction picture, in which the free dynamics of $\hat{\mathcal{O}}_f\left(t \right)$ and $\mathcal{O}_f\left(t \right)$ are corrected by the effects of the interactions determined by the quantum and classical interaction Hamiltonians (again, in the interaction picture), $\hat H_I$ and $H_I$, evolved from a reference time $t_0$ until time~$t$. 
The commutator $[\,,]$ appearing in~\eq{eq: Dyn Q} is the usual one, while the Poisson bracket of~\eq{eq: Dyn C} is defined, relative to $t_0$ and integrating in Fourier space, as:
\begin{align}
	\nonumber
	&
	\{A_f(t_1),B_f(t_2)\}_0 \equiv \sum_a \int \dd^3 \vb{k} 
	 \\ \nonumber &\times 
	\left( \frac{\partial A_f(t_1)}{\partial \Phi_{a,\vb{k}}(t_0)} \frac{\partial B_f(t_2)}{\partial \Pi_{a,\vb{k}}(t_0)} - \frac{\partial A_f(t_1)}{\partial \Pi_{a,\vb{k}}(t_0)} \frac{\partial B_f(t_2)}{\partial \Phi_{a,\vb{k}}(t_0)} \right) \,.
\end{align}
Given a pair of complex solutions to the free equations of motion,  $s_{a,k}(t)$ and $s_{a,k}^*(t)$, the classical and quantum free fields can be written as
\begin{align}
	&\Phi_{a,\vb{k}}^f(t) = s_{a,k}(t) \alpha_{a,\vb{k}} + s_{a,k}^*(t) \alpha_{a,-\vb{k}}^* \,, \label{eq: Sol Cl} \\
		&\hat\Phi_{a,\vb{k}}^f(t) = s_{a,k}(t) \hat a_{a,\vb{k}} + s_{a,k}^*(t) \hat a_{a,-\vb{k}}^\dagger \,. \label{eq: Sol Q}
\end{align}
The constants $\alpha_{a,\vb{k}}$ are obtained by imposing that the free fields $\Phi_{a,\vb{k}}^f$ must satisfy the initial conditions of $\Phi_{a,\vb{k}}$ (and likewise for their respective conjugate momenta) at time $t_0$:
	$\alpha_{a, \vb{k}} = \mathcal{W}_{a,k}^{-1}\left({\Phi_{a,\vb{k}}(t_0) \dot{s}_{a,k}^*(t_0)-\Pi_{a,-\vb{k}}(t_0) s_{a,k}^*(t_0)}\right)$,
where $\mathcal{W}_{a,k} \equiv s_{a,k}(t_0)\dot{s}_{a,k}^*(t_0) - s^*_{a,k}(t_0)\dot{s}_{a,k}(t_0) $ is the Wronskian and dots denote derivatives with respect to time, $t$.
Then, $\{\Phi_{a,\vb{k}}(t_0),\Pi_{b,\vb{p}}(t_0)\}_0=\delta_{ab}\,\delta(\vb{k}-\vb{p})$ implies that $\{\alpha_{a,\vb{k}}(t_0),\alpha^*_{b,\vb{p}}(t_0)\}_0=\mathcal{W}_{a,k}^{-1}\,\delta_{ab}\,\delta(\vb{k}-\vb{p})$. Analogously, imposing standard commutation rules among the free fields, $\comm{\hat{\Phi}^f_{a,\vb{k}}(t)}{\hat{\Pi}^f_{b,\vb{p}}(t)} = i \hbar\,  \delta_{ab} \, \delta(\vb{k} - \vb{p})$, we get
	$\comm{\hat a_{a,\vb{k}}}{\hat a_{b,\vb{p}}^\dagger} = i \hbar\,\mathcal{W}_{a,k}^{-1}\, \delta_{ab}\,\delta(\vb{k} - \vb{p})$.
Therefore, the standard commutator for annihilation and creation operators is recovered with the normalization $\mathcal{W}_{a,k}=i\hbar$, which also fixes the corresponding Poisson bracket. In this way, both classical and quantum dynamics are defined. The time evolutions \eq{eq: Dyn Q} and \eq{eq: Dyn C} have formally analogous expressions, with a key difference being that in the quantum case the order of the fields matters. Besides, whereas $\mathcal{O}(t)$ is determined given the classical initial conditions $\Phi_{a,\vb{k}}(t_0)$ and $\Pi_{a,\vb{k}}(t_0)$, $\hat{\mathcal{O}}(t)$ is only observable through its expectation value over the quantum state of the system. Since we want the classical and quantum evolutions to be as close as possible, evolving from $t_0$, we impose that classical and quantum statistics agree at $t_0$.

In inflation, we are usually interested in the quantum state being the vacuum $\ket{\Omega}$, so that any observable takes the form $\expval{\hat{\mathcal{O}}(t)} \equiv \bra{\Omega} \hat{F}^\dagger(t,t_0) \hat{\mathcal{O}}_f(t) \hat{F}(t,t_0) \ket{\Omega}$ \cite{Weinberg:2005vy}.
In a system with interactions, $\ket{\Omega}$  does not coincide with the free vacuum $\ket{0}$, which is the one that satisfies $\hat a_{a,\vb{k}} \ket{0} = 0$ (i.e.~$\ket{0}$ is an instantaneous eigenstate of the free Hamiltonian $\hat H_0$, but $\ket{\Omega}$ is not, unless the theory is free). Assuming that the system becomes asymptotically free in the limit $t_0 \to - \infty$, we can project $\ket{\Omega}$ onto the Bunch-Davies free vacuum $\ket{0}$ with the $i \epsilon$ prescription:
\begin{equation} \label{eq: in-in}
	\left<\hat{\mathcal{O}}(t) \right> = \bra{0} \hat{F}^\dagger(t,-\infty_+) \hat{\mathcal{O}}_f(t) \hat{F}(t,-\infty_-) \ket{0}\vline_{\textrm{n.b.}} \,,
\end{equation}
where n.b.\ denotes the removal of bubble diagrams and $t_\pm \equiv t(1\pm i \epsilon)$. 

In the classical counterpart, we choose stochastic initial conditions that match the quantum ones at a finite time. This means that only statistical averages will be accessible: $\expval{\mathcal{O}(t)}^{\rm Cl} \equiv \expval{F(t,t_0) \mathcal{O}_f(t)}^{\rm Cl}$,
where $\expval{\,\,}^{\rm Cl}$ denotes the average over the probability distribution function of the object inside. 
This quantity depends on the initial conditions, which we impose to be determined by quantum statistics at $t_0$ according to Eq.~\eq{eq: in-in}.

It has been common to study the classical evolution with Gaussian initial conditions, both in cases where the dynamics of inhomogeneities were treated perturbatively  --see e.g.~\cite{Green:2020whw,Ghosh:2022cny}-- and non-perturbatively, in particular in lattice calculations \cite{Caravano:2021pgc,Caravano:2024tlp,Caravano:2024moy}. The prescription above is preferable, as it exactly reproduces the quantum statistics at $t_0$. However, as we will see with two examples, this prescription does not capture the complete quantum properties through classical evolution.

\section{Scalar Bispectrum}

Let us consider the same action for the primordial curvature perturbation $\zeta$ that was used in  \cite{Green:2020whw}:
\begin{equation}\label{ec:action_scalar}
	S = \int \dd \tau \, \dd^3 \vb{x} \, a^2 M_P^2\, \epsilon \left( \zeta'^2 - \partial \zeta^2 + \frac{\lambda}{3!} \frac{\zeta'^3}{ a H}  \right) \,,
\end{equation}  
where primes denote derivatives with respect to conformal time, $\tau$, and $\lambda$ is a constant. 
For simplicity, and following \cite{Green:2020whw}, we assume a de~Sitter background metric: $a(\tau) = -1/(H \tau)$, neglecting $\order{\epsilon}$ slow-roll corrections, where $\epsilon\equiv -H'/aH^2$. The free equation of motion has $s_k(\tau)=[H^2\hbar/(4M_P^2\epsilon k^3)]^{1/2}(1+ik\tau)e^{-ik\tau}$ as a solution,
where we have imposed Bunch-Davies initial conditions at very early times and used a normalization that ensures canonical commutation relations for the creation and annihilation operators associated to the canonically normalized field $\sqrt{2\epsilon}aM_P\zeta$, being $M_P$ the reduced Planck mass.

The (quantum) bispectrum in terms of the three-point expectation value of $\hat{\zeta}_{\vb{k}}(\tau)$ reads 
$(2\pi)^{3/2}\expval{\prod_{i=1}^3\hat{\zeta}_{\vb{k}_i}(\tau)} \equiv {\delta(\vb{k}_1+\vb{k}_2+\vb{k}_3)}\,B^Q_\zeta(\tau;\{k_i\})$,
which is computed using Eq.~\eq{eq: in-in}. Hereafter, we suppress the explicit dependence of $B_\zeta^Q(\tau;\{k_i\})$ for notational simplicity. At linear order in $\lambda$, it is: 
\begin{align} \label{quantumbi}
&B^Q_\zeta =- \frac{ 2M_P^2\epsilon\lambda}{H \hbar}
  \Im{  \int_{-\infty_-}^\tau \!\!\!\!\dd \tau'\,a \prod_{i = 1}^3 s_{k_i}(\tau) {s'^*_{k_i}}(\tau')} \,,\\  
 &\lim_{\tau \to 0}B^Q_\zeta =  \frac{ \lambda \, \hbar^2 H^4}{16 \epsilon^2 M_P^4 } \frac{1}{k_1 k_2 k_3 (k_1+k_2+k_3)^3} \,. \label{quantumbiLT}
\end{align}
Taking into account that $s_k\sim \sqrt{\hbar}$, the quantum bispectrum scales as $\hbar^2$, consistently with a purely perturbative calculation. The fact that the leading order is suppressed by $\hbar$ (squared) is a consequence of the quantum nature of the vacuum, with no classical analogue. 

Considering $\zeta$ to be instead a classical field, its time evolution at $\order{\lambda}$ is given by Eq.~\eqref{eq: Dyn C}:
\begin{align} \nonumber
	& \zeta_{\vb{k}_1}(\tau) = \zeta_{\vb{k}_1}^f(\tau)  - \frac{M_P^2\epsilon\lambda}{H\hbar} \int_{\tau_0}^\tau \dd \tau' a(\tau') \\ 
	&\times \int \frac{\dd^3 \vb{p} }{(2\pi)^{3/2}}   \Im{s_{k_1}(\tau) s'^*_{k_1}(\tau')}   
	 \left.{{\zeta^f}'_{\vb{p}}}{{\zeta^f}'_{\vb{k}-\vb{p}}}\right|_{\tau'}  \,.\label{eq: Part Case B Poisson}
\end{align}
From \eq{eq: Part Case B Poisson}, we obtain the classical connected three-point function of $\zeta_{\vb k}$, noting that $\langle\zeta_{\vb{k}_1}^f\zeta_{\vb{k}_2}^f \zeta_{\vb{k}_3}^f\rangle^{\rm Cl}$ is non-vanishing and necessary to match the quantum and classical bispectra at $\tau_0$. To perform the calculation we  need the two- and three-point statistical averages of the coefficients $\alpha_{\vb k}$, see Eq.~\eq{eq: Sol Cl}. The two-point statistics of $\alpha_{\vb{k}}$ at $\order{\lambda^0}$ is determined by imposing equality of the quantum and classical tree-level power spectra at $\tau=\tau_0$. This leads to $\expval{\alpha_{\vb{k}_1} \alpha_{\vb{k}_2}} = 0 \quad {\rm and} \quad \expval{\alpha_{\vb{k}_1} \alpha^*_{\vb{k}_2}} = \delta(\vb{k}_1 - \vb{k}_2)/2$, which determine the contribution generated by a single insertion of $H_I$ at $\mathcal{O}(\lambda)$. The three-point statistic of $\alpha_{\vb{k}}$ at $\order{\lambda}$ accounts for the non-vanishing initial bispectrum, and is determined by requiring the classical bispectrum to be equal to the quantum one at $\tau_0$.
The only independent non-vanishing three-point classical average of $\alpha_{\vb{k}}$ and $\alpha ^*_{\vb{k}}$ is
\begin{align} \nonumber
\expval{\alpha_{\vb{k}_1} \alpha_{\vb{k}_2} \alpha_{\vb{k}_3}}^{\rm Cl} =  \frac{iM_P^2\epsilon \lambda }{H \hbar}\frac{\delta\left(\sum\limits_{i=1}^3\vb{k}_i\right)}{(2 \pi)^{3/2}}  \int_{-\infty_-}^{\tau_0} \!\!\!\dd \tau' a \prod_{i=1}^3s'^*_{k_i}\,. 
\end{align}
Subtracting the classical bispectrum from its quantum counterpart we obtain:
\begin{align}
\label{eq: Delta B} 
	\Delta B_\zeta = -\frac{ \hbar^2 M_P^2\epsilon\lambda }{ 4  H } \int_{\tau_0}^\tau \!\!\dd \tau'a \prod_{i = 1}^3 \partial_{\tau'} g_{k_i}(\tau,\tau')\,,
\end{align}
where $ g_k(\tau,\tau')= 2 \hbar^{-1}\Im{s_{k}(\tau) s_{k}^*(\tau')}$ is the Green function, see also \cite{Green:2020whw}. The difference \eq{eq: Delta B} vanishes for $k\tau_0\to 0$. Thus, for classical and quantum evolution to coincide at all times, the initial matching between the two needs to be imposed choosing $\tau_0$ such that $\zeta$ is already constant. That would be a self-defeating strategy, as it requires doing the complete quantum calculation until the modes cross the horizon and freeze. Only in that case $\Delta B_\zeta(\tau;\{k_i\})$ will vanish at later times. This conclusion appears consistent with the notion that $\zeta$ classicalizes as it freezes at comoving distances sufficiently larger than $1/(a H)$; see \cite{Launay:2024trh} for a related discussion. 

Performing the time integral in \eq{eq: Delta B} and taking the limit $\tau\to 0$, we obtain
\begin{align}
\label{eq: B dif Q vs C} \nonumber
& \lim_{\tau\to0}\Delta B_\zeta = \frac{\lambda\, H^4 \hbar^2}{4 \epsilon^2M_P^4} \Re\Bigg\{f_0 - \bigg[ f_3 \\ 
&+ \frac{3 k_3^4 (k_1^2 + k_2^2 - k_3^2) - 14/3 \,k_1^2 k_2^2 k_3^2}{K_0^3 K_1^3 K_2^3 K_3^3} + {\rm perms.} \bigg] \Bigg\} \,,
\end{align}
where we define $K_0 = \sum_i ^3 k_i$, $K_j = K_0 - 2k_j$, $j={1,2,3}$; and also $f_i={e^{i K_i \tau_0} (2-2i K_i \tau_0-( K_i \tau_0)^2)}/{32 k_1 k_2 k_3 K_i^3}$, $i={0,1,2,3}$.
The consistency relation between the power spectrum and the squeezed limit of the bispectrum \cite{Maldacena:2002vr, Creminelli:2004yq} is satisfied for \eqref{quantumbiLT} and \eqref{eq: B dif Q vs C}; therefore, it also holds for the classical bispectrum.
In the quasi-de Sitter approximation we are using, the consistency relation reduces simply to $p^3 k^3 \, B_\zeta(\tau \to 0;p,k,k) \to 0 $ as $p\to 0$. 
In the limit $p\ll k$, $f_0$ scales as $p^{-1}$, and is  thus compatible with the consistency relation. The other $f_i$ of \eqref{eq: B dif Q vs C} include denominators which, a priori, may seem problematic in the $p\to 0$ limit due to their $p^{-4}$ scaling. However, $\Delta B_\zeta(\tau\rightarrow 0;\{k_i\})$ scales as $p^0$ in the squeezed limit due to a subtle cancellation of independently divergent terms, therefore fulfilling the consistency relation, contrary to what the classical result in~\cite{Green:2020whw} implies. 
Similarly, a superficial look at \eqref{eq: B dif Q vs C} may give the impression of divergences (poles) arising in the limit in which two momenta are collinear (i.e. a folded momentum configuration), due to the dependence on $K_i$ in the denominator. However, it can be explicitly checked that \eqref{eq: B dif Q vs C} is not divergent in these configurations. Moreover, the bispectrum is not generically enhanced in said configurations for arbitrary values of $\tau_0$. This means that neither poles nor an enhanced signal in near-folded limits are a signature of classical evolution from a finite time onwards.

Unlike in the quantum calculation, there is no classical analogue of the $i\epsilon$ prescription that yields a well-defined $\tau_0\to -\infty$ limit while remaining compatible with the consistency relation.
Taking inspiration from the (quantum) $i\epsilon$ prescription, one could naively introduce a factor $e^{-\epsilon (\tau-\tau')}$ in the integrand of \eqref{eq: Delta B} in an attempt to remove terms linear and quadratic in $ \tau_0$ while taking $\tau_0\to -\infty$. That would lead to only the second line of~\eqref{eq: B dif Q vs C} surviving, which would not satisfy the consistency relation, and would also give rise to poles at physical momenta in the folded configurations mentioned above. {It is therefore necessary to keep the contribution from the matching time to verify the absence of poles, which is not an artefact of the choice of initial conditions. It follows from the integrals being defined over a finite time interval.} 
The classical bispectrum was computed in \cite{Green:2020whw} integrating from $\tau = -\infty$, but dropping the contribution to the integral coming from that limit --as the contribution from the end of inflation was considered there to be dominant-- leading to the appearance of poles 
\footnote{It was argued in \cite{Green:2020whw} (independently of the treatment of the $\tau_0\rightarrow - \infty$ limit) that scenarios with dissipative interactions {\cite{Berera:1995ie,Berera:1995wh,LopezNacir:2011kk}} or with excited states \cite{Holman:2007na,Flauger:2013hra} can be considered classical and lead to enhanced folded configurations. In our present work we are not concerned with these situations, {as we focus on single-field inflation.} Besides, if classical dynamics were to be associated to abundant and continuous particle production, it would appear reasonable to set a finite time from which sufficient particles are produced. We also note that the divergences found in de Sitter models with $\alpha$-vacua \cite{Ghosh:2022cny,Shukla:2016bnu} have a closely related origin: the corresponding mode functions contain both positive- and negative-frequency phases.}.

\section{Scalar Induced tensor fluctuations}

Let us now consider an inflaton field, $\phi$, minimally coupled to gravity in quasi-de Sitter spacetime. We work in the gauge in which the only scalar perturbation is the field perturbation $\delta\phi$ and the spatial part of the metric is $a^2(e^h)_{ij}$, with $\partial_i h_{ij}=0$ and $h_{ij}$ traceless, which defines the tensor modes. At lowest order in $\epsilon$, which allows to neglect shift and lapse perturbations \cite{Ballesteros:2024zdp} 
the interactions between $h_{ij}$ and $\delta\phi$ come from the (assumed standard) kinetic term of $\phi$:
\begin{align} \label{actiongrav}
	S \supset \int \dd \tau \, \dd^{3} \vb{x} \, \frac{a^{2}(\tau)}{2}
	\left( h_{ij} - \frac{1}{2} h_{ik}h_{kj}\right) \partial_i \delta\phi\, \partial_j \delta\phi \,.
\end{align} 
The one-loop correction to the (quantum) tensor power spectrum, $\mathcal{P}_h^{Q\,1l}(\tau,k)$, 
can be computed using the in-in formalism \cite{Weinberg:2005vy} and the polarization basis $\{e_{ij}^\gamma(\vb{k})\}_{\gamma=+,\times}$, in which $h_{ij}(\vb{x})=(2\pi)^{-3/2}\int \dd^3\vb{k}\,e^{i\vb{k}\cdot\vb{x}}\sum_{\gamma}e_{ij}^\gamma(\vb{k})h_{\vb{k}}^\gamma(\tau)$ \cite{Ballesteros:2024cef}. 

The procedure to determine the classical counterpart to $\mathcal{P}_h^{Q\, 1l}(\tau,k)$ at the same perturbative order is identical to the one we applied for the scalar bispectrum. First, we determine the statistics of $\alpha_{a,\vb{k}}$, see  Eq.~\eqref{eq: Sol Cl}, now for $\delta\phi$ and $h_{ij}$, matching the classical and quantum statistics at $\tau=\tau_0$. Second, we use this matching with the classical evolution encoded in \eqref{eq: Dyn C} to compute the classical power spectrum at $\tau>\tau_0$. 

The difference between the quantum and classical power spectra is, see Appendix 2:
\begin{align}\label{ec:result_GW}
& \Delta \mathcal{P}_h(\tau,k) = \frac{\hbar^2k^3}{16\pi^2}\int \frac{\dd ^3 \vb{p}}{(2 \pi)^3} p^4 \sin^4 \theta\nonumber \int_{\tau_0}^\tau  \dd \tau' a^2 \int_{\tau_0}^{\tau'} \dd \tau'' a^2 \\ & \quad\quad\quad \times g_k^{h}( \tau,\tau') g_k^{h}( \tau,\tau'') g_p^{\phi}( \tau',\tau'')g_q^{\phi}( \tau',\tau'')\,,
\end{align}
where  $g_k^{\phi,h}$ denote the Green functions for $\delta\phi_{\vb k}$ and for $h_{\vb k}^\gamma$ (with $g_k^{h}$ being independent of the polarization). The angle between $\vb{k}$ and~$\vb{p}$ is denoted by $\theta$, and $\vb{q}\equiv\vb{k}-\vb{p}$. 

In the limit $k\tau_0\rightarrow 0$, the tensor modes become constant, $g_k^{h}( \tau,\tau')\rightarrow 0$ and $\Delta \mathcal{P}_h(\tau,k)$ vanishes. 
Away from this trivial (non-dynamical) limit, it is different from zero. We can obtain the  behaviour for $k\tau_0\rightarrow -\infty$
with the sub-horizon limits: $s_k ^\phi = \sqrt{\hbar}\,e^{-ik\tau}/a\,\sqrt{2k}$ and ${s}_k^h = 2{s}_k ^\phi/{a\,M_P}$. Thanks to the suppression of the Green functions in this limit, the super-horizon contribution of scalar modes running in the loop is negligible, which justifies using the previous expression for ${s}_k ^\phi$. Finally, we obtain $\Delta \mathcal{P}_h(\tau,k)\propto (k\tau_0)^2$ for $k\tau_0\rightarrow -\infty$.

\section{Discussion}

Differences between quantum and classical computations of cosmological correlation functions are unveiled only if interactions are taken into account. At tree-level, it is necessary to consider correlations of three or more fields. At loop-level, the power spectrum is sufficient.
These differences are controlled by products of Green functions, i.e.~by products of imaginary parts. Eqs.~\eq{eq: Delta B} and \eq{ec:result_GW} show this explicitly for the two examples we have considered. This dependence is not an accident: classical and quantum dynamics have analogous functional forms, see Eqs.~\eq{eq: Dyn Q} and \eq{eq: Dyn C}, and the mismatch comes from the nature of the quantum building blocks, which are operators whose ordering
matters. Indeed, for any two operators $\hat A$ and $\hat B$ evaluated at different times: $\hat A(\tau_1)\hat B(\tau_2) = \frac{1}{2}\left[\hat A(\tau_1),\hat B(\tau_2)\right]+\frac{1}{2}\left(\hat A(\tau_1)\hat B(\tau_2)+\hat B(\tau_2)\hat A(\tau_1)\right)$. 
The anti-commutator is insensitive to operator ordering and is therefore classically reproducible once
the initial statistics are matched. However, the commutator is intrinsically quantum and it
translates into products of Green functions in explicit computations, which implies that its effect cannot be reproduced in general with classical dynamics. A necessary condition for classical dynamics to be a good approximation is therefore $\Re{s_k(\tau)s_k^*(\tau')}\gg \Im{s_k(\tau)s_k^*(\tau')}$ --compare e.g.~Eqs.~\eq{quantumbi} and \eq{eq: Delta B}-- which means that the anti-commutator dominates over the commutator, see also~\cite{Launay:2024trh}. 

However, even if that condition is met in a classical calculation, there is always a residual dependence on the time, $\tau_0$, at which the matching between classical and quantum initial conditions is imposed. After performing the time integrations in Eq.~\eq{eq: Dyn C}, products of Green functions lead to a dependence on $\tau_0$ that is quadratic in the examples we have explored. This implies an exponential dependence on the number of e-folds ($N=\int a H\, d\tau $) elapsed between $\tau_0$ and the time of horizon crossing. 

We can anticipate the residual $\tau_0$ dependence of any classical observable inspecting the time scaling of the 
interaction vertices. In the sub-horizon regime, for a fluctuation $\Phi$ with a standard kinetic term,
the modes go as $s_k^\Phi\propto e^{-ik\tau}/a$. Therefore, an interaction of the schematic form $\sqrt{-g}\,\partial^n\Phi^m/a^n$, where $\partial$ denotes spatial or conformal-time derivatives, scales as $a^{4-n-m}$, which diverges at early times unless $4\geq n+m$. The quantum calculation is insensitive to this scaling because the $i\epsilon$ prescription exponentially suppresses it. In the classical case, by contrast, the absence of such a prescription implies that the vertex scales as $\tau_0^{n+m-4}$. This dependence is essential for the consistency of classical predictions, as we discussed for the scalar bispectrum. 

The accumulated error from using classical dynamics to compute the equilateral configuration of the scalar bispectrum for the action of Eq.~\eq{ec:action_scalar} in quasi-de Sitter can be evaluated analytically:
${\Delta B}/{B^{Q}} = 20 + \frac{1}{8}\textrm{Re}\bigg\{81 e^{-i z} (z^2-2z i-2)-e^{-3 i z} \left(9 z^2-6 i z-2\right) \bigg\}$,
where $z\equiv -k\tau_0 = k/a(\tau_0)H=e^{\Delta N}$, being $\Delta N$ the number of e-folds elapsed from $\tau_0$ until the time at which $k=aH$. If the matching is done one e-fold before $k=aH$, then $\abs{\Delta B/B^{Q}}\sim 50$. The issue can be illustrated as well using the (one-loop) tensor power spectrum induced by scalar fluctuations, $\mathcal{P}_h^{Q\,1l}$. From our results, we expect  ${\Delta \mathcal{P}_h}/{\mathcal{P}_h^{Q\, 1l}}\sim e^{2\,\Delta N}$ to capture the order of magnitude of this ratio, which again indicates an exponential discrepancy between classical and quantum calculations.

No matter how well the classical initial conditions are set to mimic the quantum dynamics, the classical evolution rapidly departs from the quantum result. 
The root of this behaviour is that the classical limit, $\hbar \rightarrow 0$, is not appropriate: the fluctuations originate from the quantum vacuum and must be computed using an intrinsically quantum method. Whenever deviations from the free Gaussian theory are probed, classical dynamics can dramatically fail to capture the quantum nature of the fluctuations. Due to the dependence on $\tau_0$, great care is needed to interpret predictions for inflationary statistics obtained with classical methods: not only perturbative (classical Green functions) but also non-perturbative ones, specifically lattice simulations (whose use is rather recent and has mainly been focused on the scalar power spectrum). The dependence on $\tau_0$ can be hidden if the interactions are very small \cite{Caravano:2021pgc} or if the integrals are dominated by time intervals away from $\tau_0$, where the interactions are most relevant and the modes can be enhanced~{\cite{Jamieson:2025ngu,Caravano:2024moy}.} However, this does not ensure the accuracy of the classical calculation. 

For instance, reference \cite{Caravano:2024moy} studied transient ultra slow-roll inflation in a regime that produces a large spectral peak at specific comoving scales, which can lead to the production of primordial black holes in the radiation or matter epochs. One can verify, either with a concrete inflationary potential \cite{Ballesteros:2017fsr} or with an ad-hoc parametrization of the dynamics \cite{Ballesteros:2024zdp}, that the proxy condition $\Re{s_k(\tau)s_k^*(\tau')}\gg \Im{s_k(\tau)s_k^*(\tau')}$ is not satisfied for the modes of interest to describe the peak of the power spectrum at the times around the transitions associated with the ultra slow-roll stage, where the interactions are more important 
\footnote{The situation can differ for initially excited (non-Bunch-Davies) states, for which the real part may be enhanced while the imaginary part is fixed by the commutator, equivalently by the Green’s function, and is therefore insensitive to the particular excitation.}.
The same can be expected to occur in other scenarios that feature super-horizon evolution of the modes. This suggests that the application of lattice methods in such cases is hindered.
 
We can consider the extreme case in which the interactions are strictly localized in time. An example is provided by the tree-level bispectrum for the action of Eq.~\eq{ec:action_scalar} with a toy interaction in which $\lambda(\tau) \propto \delta(\tau - \tau_*)$.  At late times and in the equilateral limit,
${\Delta B}/{B^{\rm Q}} = {\sin ^2(k \tau_*)}/{(2 \cos (2 k \tau_*)+1)}.
$
This ratio diverges for $2k\tau_*= {2\pi}/{3}-2\pi\,c$, with $c\in\mathbb{N}$. While the divergence in $k$ is an artefact of the instantaneous interaction, the example illustrates that localized interactions {can screen the sensitivity to $\tau_0$, which does not imply that the classical and quantum predictions agree.}

We conclude by stressing that, in our classical calculation starting from a finite initial time, the scalar bispectrum does not develop poles in folded configurations. Moreover, its squeezed limit satisfies the consistency relation. Therefore, poles in this observable are not a signal of classical evolution in this setup. We speculate that, analogously, poles will be absent from higher-order correlators when the evolution is defined over a finite time interval.
Our analysis is not conclusive about the appearance of poles --or enhanced bispectra-- in folded configurations as a consequence of classical evolution from $\tau = -\infty$, see \cite{Green:2020whw}. However, the lack of a robust analogue to the $i \epsilon$ prescription actually implies a general impediment for reaching a conclusion in that regard.

{\bf Acknowledgements.} We thank Nima Arkani-Hamed, Daniel Baumann, Daniel Green, Yoann Launay, Rafael Porto, Gerasimos Rigopoulos and Paul Shellard for comments and discussions.  This work has been funded by the grants: 1)  PID2021-124704NB-I00 funded by MCIN/AEI/10.13039/501100011033 and by ERDF A way of making Europe, 2)
CNS2022-135613 funded by MICIU/AEI/10.13039/501100011033 and by the European Union NextGenerationEU/PRTR, and 3)
Centro de Excelencia Severo Ochoa CEX2020-001007-S funded by MCIN/AEI/10.13039/501100011033. JGE is
supported by a PhD contract {\it contrato predoctoral para formación de doctores} (PRE2021-100714) associated to
the Severo Ochoa grant: CEX2020-001007-S-21-3. APR has been supported by a NASA ATP award:
80NSSC24K0937. 

\onecolumngrid

\appendix

\section{Appendix 1. Classical and quantum dynamics} \label{app: QvsC din}

We will derive here the equations governing both classical and quantum field dynamics, Eqs.~(1) and~(2). We will do so
by expanding the full field dynamics perturbatively over the free (non-interacting) case, which is known as the
interaction picture.
\subsection{Classical dynamics}
Beginning with the classical case, consider a system of coupled classical particles described by the action \footnote{The generalization to fields follows from taking into account that, in Fourier space, fields correspond to a set of coupled particles labelled with the continuous index $\vb{k}$ instead of the discrete one $a$.}
\begin{equation}
            S = \int \dd t \, L(q_a(t),\dot{q}_a(t);t) \,,
\end{equation}
where $a$ is the index that characterizes the different particles. However, we will omit the explicit dependence on
the label $a$ when we refer to the dependence on the set $\{q_a\}$. As it is well known, the corresponding set of Euler-Lagrange equations follows from varying the action,
\begin{equation}\label{ec:EL}
            \frac{\partial L}{\partial q_a} - \frac{\dd}{\dd t} \frac{\partial L}{\partial \dot{q_a}} = 0\,,
\end{equation}
and the corresponding Hamiltonian follows from Legendre transforming the Lagrangian,
\begin{equation}\label{ec:def_hamiltonian}
            H(q(t),p(t);t) = \sum_a\left( p_a\, \dot{q}_a(q,p) \right) - L(q,\dot{q}(q,p);t)\,, \quad {\rm where} \quad p_a \equiv \frac{\partial L}{\partial \dot{q}_a} \,.
\end{equation}
Combining \eqref{ec:EL} and \eqref{ec:def_hamiltonian}, one obtains Hamilton's equations of motion,
\begin{equation}\label{ec:hamilton_eoms}
            \dot{q}_a = \frac{\partial H}{\partial p_a} \quad {\rm and} \quad \dot{p}_a = - \frac{\partial H}{\partial q_a} \,,
\end{equation}
where we have used that
\begin{equation}
            \frac{\partial p_a}{\partial p_b} = \frac{\partial q_a}{\partial q_b} = \delta_{ab} \quad {\rm and} \quad \frac{\partial q_a}{\partial p_b} = \frac{\partial p_a}{\partial q_b} = 0\,.
\end{equation}
Consider a generic function $\mathcal{O}(p,q)$ that does not depend on time explicitly. Its time derivative follows from \eqref{ec:hamilton_eoms} \footnote{
If $\mathcal{O}$ depended explicitly on time, an extra term $\partial \mathcal{O}/\partial t$ would be added on the right-hand side.
},
\begin{equation}\label{ec:poisson_brackets}
            \frac{\dd}{\dd t} \mathcal{O}(q(t),p(t)) = \sum_a \left( \frac{\partial \mathcal{O}}{\partial q_a} \dot{q}_a + \frac{\partial \mathcal{O}}{\partial p_a} \dot{p}_a \right) = \sum_a \left( \frac{\partial \mathcal{O}}{\partial q_a} \frac{\partial H}{\partial p_a}-\frac{\partial \mathcal{O}}{\partial p_a} \frac{\partial H}{\partial q_a} \right) \equiv \{\mathcal{O},H\} \,,
\end{equation}
where in the last step we have introduced the Poisson brackets, satisfying
\begin{equation}
\{q_a(t),p_b(t)\} = \delta_{ab} \quad \text{and} \quad\{q_a(t),q_b(t)\} = \{p_a(t),p_b(t)\} = 0\,.
\end{equation}
Using the Poisson brackets, Hamilton's equations can then be rewritten in a compact manner as
\begin{equation}
            \dot{q}_a(t) = \{q_a,H\} \quad {\rm and} \quad \dot{p}_a(t) = \{p_a,H\} \,.
\end{equation}
 
We now consider a generic function $\mathcal{O}(t)$ of the form $\mathcal{O}(t)=\mathcal{O}(q(t),p(t))$, i.e.\ which solely depends on time implicitly through $p$ and $q$. We formally express it as the time evolution of its initial value by means of an operator (acting on the variables $q_a^0$ and $p_a^0$) of the form
\begin{equation}\label{ec:evol_o}
            \mathcal{O}(t)=\mathcal{O}(q(t),p(t)) = U(t,t_0) \mathcal{O}(q^0,p^0) \,.
\end{equation}
The dynamics of $\mathcal{O}(t)$ is therefore encoded in the {\it time evolution operator} $U(t,t_0)$, which satisfies
\begin{align}
            \dot{U}(t,t_0) \mathcal{O}_0 = \{U(t,t_0) \mathcal{O}_0 , U(t,t_0) H(q^0,p^0;t)\}  \quad \textrm{and then} \quad
            \dot{U}(t,t_0) \, \cdot \,= \{U(t,t_0) \, \cdot \, , U(t,t_0) H(q^0,p^0;t)\} \,,
\end{align}
where we denote $\mathcal{O}_0 \equiv \mathcal{O}(q^0,p^0)$.
 
Solving Hamilton's equations, and therefore computing $U(t,t_0)$, is unfeasible except for a small set of Hamiltonians. One can, however, construct perturbative solutions around one of the few solvable cases, the free one. Splitting the Hamiltonian into its free (and quadratic) contribution $H_f$ and its interaction contribution $H_{\rm int}$, such that $H = H_f + H_{\rm int}$, the solution for a general function built from the free positions and momenta can be constructed as:
\begin{equation}\label{ec:evol_of}
            \mathcal{O}_f(t) = U_f(t,t_0) \mathcal{O}_0 \quad {\rm with} \quad \dot{U}_f(t,t_0) \, \cdot \,= \{U_f(t,t_0) \, \cdot \, , U_f(t,t_0) H_f(q^0,p^0;t)\}_f \,,
\end{equation}
where the Poisson bracket $\{\,,\}_f$ denotes Eq.~(\ref{ec:poisson_brackets}) specialized to the case in which the derivatives are taken with respect to the free position and momentum variables.
Here, we have defined $\mathcal{O}_f(t)\equiv \mathcal{O}(q_f(t),p_f(t))$, where $q_f$ and $p_f$ are solutions to the equations of motion derived from $H_f$. Let us now introduce an operator $F(t,t_0)$ that relates functions of position and momentum computed in the full and free theories \footnote{Formally, such operator can be related to the time evolution operators in the full and free theories as $U(t,t_0) U_f^{-1}(t,t_0) \equiv F(t,t_0)$, as it follows from \eqref{ec:evol_of} and \eqref{ec:evol_o}.},
\begin{equation}\label{ec:def_f}
            \mathcal{O}(t) \equiv F(t,t_0)\mathcal{O}_f(t) \,.
\end{equation}
Differentiating \eqref{ec:def_f} with respect to time,
\begin{align}
	\nonumber
\dot{\mathcal{O}}(t) & =\dot{F}(t,t_0)\mathcal{O}_f(t) + F(t,t_0)\dot{\mathcal{O}}_f(t)
=\dot{F}(t,t_0) \mathcal{O}_f(t) + F(t,t_0) \{\mathcal{O}(q_f(t),p_f(t)),H_f(q_f(t),p_f(t);t)\}_f\\ & =\dot{F}(t,t_0)\mathcal{O}_f(t)+\{\mathcal{O}(q(t),p(t)),H_f(q(t),p(t);t)\}\,,
\end{align}
where in the last step we have used that, when acting on any function of the free variables, $F(t,t_0)$ replaces $q_f\to q$, $p_f\to p$ without altering the functional form of the expression. Using the equation of motion for $\dot{\mathcal{O}}(t)$ we get
\begin{equation}
\dot{F}(t,t_0)\mathcal{O}_f(t) =\{\mathcal{O}(t), H(q(t),p(t);t)-H_f(q(t),p(t);t)\}=F(t,t_0)\{\mathcal{O}_f(t), H_{\rm int}(q_f(t),p_f(t);t)\}_f \,.
\end{equation}

In order to formulate classical mechanics in a way parallel to quantum mechanics, we seek a perturbative solution to the dynamical equation for $F(t,t_0)$. To this end, we must first define the different-time Poisson brackets acting on free variables (see also~\cite{Nikolic:1998tp,deAlwis:2015ioa}).
Introducing a reference time $t_0$ (which corresponds to the time at which the initial conditions of the system are set), we define
\begin{align}\label{ec:diftime_poisson}
	\nonumber
	&\{A(q_f(t_1),p_f(t_1)),B(q_f(t_2),p_f(t_2))\}_0 \equiv \sum_a\frac{\partial A_f(t_1)}{\partial q_{a}^0} \frac{\partial B_f(t_2)}{\partial p_{a}^0} - \frac{\partial A_f(t_1)}{\partial p_{a}^0} \frac{\partial B_f(t_2)}{\partial q_{a}^0} \\
	& \qquad = \sum_{a,b}\left(\frac{\partial A_f(t_1)}{\partial q_{f,a}(t_1)} \, \frac{\partial A_f(t_1)}{\partial p_{f,a}(t_1)} \right)
	\begin{pmatrix}
		\{q_{f,a}(t_1),q_{f,b}(t_2)\}_0 \quad \{q_{f,a}(t_1),p_{f,b}(t_2)\}_0 \\[6pt]
		\{p_{f,a}(t_1),q_{f,b}(t_2)\}_0 \quad \{p_{f,a}(t_1),p_{f,b}(t_2)\}_0
	\end{pmatrix}
	\begin{pmatrix}
		\frac{\partial B_f(t_2)}{\partial q_{f,b}(t_2)} \\[10pt]
		\frac{\partial B_f(t_2)}{\partial p_{f,b}(t_2)}
	\end{pmatrix}\,,
\end{align}
where $q_{a}^0 \equiv q_{a}(t_0)$ and $p_{a}^0 \equiv p_{a}(t_0)$. The Poisson bracket $\{\,,\}_0$ reduces to $\{\,,\}_f$ when evaluated at equal times, $t_1 = t_2$, provided that $\{q_{f,a}(t),p_{f,b}(t)\}_0 = \delta_{ab}$ \footnote{This is the case for a free Hamiltonian $H_f=f_p(t) p_f^2+f_q(t) q_f^2$ (we write it for a single particle for simplicity). Indeed, in this case the Euler-Lagrange equations are linear, and therefore a general solution can be written as a linear combination of two independent solutions; in particular, a complex solution and its complex conjugate. Then, we can write the general solution for $q_f$ analogously to Eq.~(3), rewrite the coefficients of the linear combination of solutions (in Eq.~(3), $\alpha$ and $\alpha^*$) in terms of $q^0$ and $p^0$, and check explicitly that $\{q_f(t),p_f(t)\}_0 =1$.}, since $\{q_{f,a}(t),q_{f,b}(t)\}_0 = \{p_{f,a}(t),p_{f,b}(t)\}_0 = 0$ by construction.

We can solve the equation for $F(t,t_0)$ perturbatively, obtaining
\begin{equation}\label{ec:Fclass}
            F(t,t_0) \, \cdot \,= 1 + \sum_{n = 1} \int_{t_0}^t \dd t_1 \int_{t_0}^{t_1} \dd t_2 \cdots \int_{t_0}^{t_{n-1}} \dd t_n \, \{ \cdots\{\{\, \cdot \, , H_I(t_1)\}_0, H_I(t_2)\}_0 \cdots , H_I(t_n)\}_0 \,,
\end{equation}
where we have defined $H_I(t) \equiv H_{\rm int}(q_f(t),p_f(t);t)$.
With this equation, it can be shown perturbatively that $F(t,t_0)A_f(t)B_f(t) = \left( F(t,t_0)A_f(t)\right) \left( F(t,t_0)B_f(t)\right)$ is satisfied order by order, as would be expected.

\subsection{\bf Quantum dynamics}

The quantum mechanical description of a system can be formally understood as the promotion of the Poisson brackets to commutators, $\{\,,\}_0 \to \frac{1}{i \hbar} \left[\,,\right]$, and the corresponding promotion of magnitudes to operators, $O(t) \to \hat{O}(t)$ --although for simplicity we will not write the hat. Thus, the dynamics of an operator \footnote{
By assuming time evolution lies on the operators rather than the states, we are working in the Heisenberg picture, as usually done in cosmology.
}
which functionally depends on the position and momentum operators is given by
\begin{equation}
            \frac{\dd}{\dd t} \mathcal{O}(q(t),p(t)) = \frac{1}{i \hbar} \comm{\mathcal{O}(q(t),p(t))}{H(q(t),p(t);t)} \,.
\end{equation}
Analogously to the previous classical discussion, we define a time evolution operator such that any time dependent operator $\mathcal{O}(t)$ ($\equiv \mathcal{O}(q(t),p(t))$, i.e. which only depends on time implicitly, as in the classical case discussed above) evolves as
\begin{equation}
            \mathcal{O}(t) = U^{-1}(t,t_0) \mathcal{O}_0 U(t,t_0)\,, \quad {\rm where} \quad \dot{U}(t,t_0) = \frac{1}{i \hbar} U(t,t_0) H(q(t),p(t);t) \,.
\end{equation}
The free version of the operator $\mathcal{O}(t)$, comprised of $q_f$ and $p_f$ operators which are solutions to the free Hamiltonian, evolves as
\begin{equation}
            \mathcal{O}_f(t) = U_f^{-1}(t,t_0) \mathcal{O}_0 U_f(t,t_0)\,, \quad {\rm where} \quad \dot{U}_f(t,t_0) = \frac{1}{i \hbar} U_f(t,t_0) H_f(q_f(t),p_f(t);t) \,.
\end{equation}
Expanding around the free solution,
\begin{equation}
            \mathcal{O}(t) = F^{-1}(t,t_0) \mathcal{O}_f(t)F(t,t_0)\,,\quad {\rm where} \quad \dot{F}(t,t_0) = \frac{1}{i \hbar} H_I(t) F(t,t_0) \,.
\end{equation}
The definition of $H_I(t)$ is the natural quantum extension of the one in the classical case: the functional form of the interaction Hamiltonian evaluated in the position and momentum operators that solve the free Hamiltonian. Again, this equation can be solved perturbatively, yielding the well-known solution
\begin{equation}
            F(t,t_0) = T\exp\left( \frac{1}{i \hbar} \int_{t_0}^t \dd t' H_I(t') \right) \,,
\end{equation}
analogous to the classical case \eqref{ec:Fclass} \footnote{In this respect, we note that the formalism discussed in \cite{Musso:2006pt} --written in terms of the Lagrangian, but equivalent to the Hamiltonian treatment adopted here, and extended to include the $i\epsilon$ prescription in \cite{Adshead:2009cb, Senatore:2009cf}-- describes interacting quantum fields in a manner formally analogous to classical field theory, which is possible since the equations of motion formally coincide. However, the formalism in \cite{Musso:2006pt} is purely quantum, as it involves (non-commuting) quantum operators rather than classical (commuting) functions.},
\begin{equation}
            F^{-1}(t,t_0) \, \cdot \, F(t,t_0) = 1 + \sum_{n = 1} \left( \frac{1}{i\hbar}\right)^n \int_{t_0}^t \dd t_1 \int_{t_0}^{t_1} \dd t_2 \cdots \int_{t_0}^{t_{n-1}} \dd t_n \,\comm{\cdots \comm{\comm{\, \cdot \, }{ H_I(t_1)}}{ H_I(t_2)}}{ \cdots , H_I(t_n)} \,.
\end{equation}
We stress that this goes beyond the in-in or in-out formalism, since we are just writing the dynamics of a generic operator in the Heisenberg picture in terms of the solution in the interaction picture.

\section{Appendix 2. Scalar-induced tensor fluctuations: classical calculation} \label{appscalind}

We consider the Fourier modes of the free fields corresponding to the components of the tensor fluctuations in the polarization basis $h_{\vb{k}}^\gamma$ and the canonically normalized scalar field $\chi_{c,\vb{k}}=a\delta\phi$. As in Eq.~(3), they can be written as
\begin{align}
	h_{\vb{k}}^{f,\gamma}(\tau) &=\alpha_{\vb{k}}^\gamma \tilde{s}_k(\tau) + \alpha_{-\vb{k}}^{\gamma,*} \tilde{s}^*_k(\tau) \,,\\
	\chi^f_{c,\vb{k}}(\tau) &= \alpha_{\vb{k}} s_{k}(\tau) + \alpha^*_{-\vb{k}} s_{k}^*(\tau) \,,
\end{align}
where $\tilde{s}_k(\tau)$ and $s_{k}(\tau)$ are the corresponding mode functions (solutions to the equation arising from the free Hamiltonian). Following Eq.~(2), the product $h_{\vb{k}}^\gamma h_{\vb{k}'}^{\gamma'}$ can be expanded in terms of such free fields to second order in the interaction Hamiltonian in the interaction picture $H_I$:
\begin{equation} \nonumber
	\left.h_{\vb{k}}^\gamma h_{\vb{k}'}^{\gamma'}\right|_{\tau} =  \left.h_{\vb{k}}^{f,\gamma} h_{\vb{k}'}^{f,\gamma'}\right|_{\tau}  +\int_{\tau_0}^\tau \dd \tau' \{h_{\vb{k}}^{f,\gamma}(\tau) h_{\vb{k}'}^{f,\gamma'}(\tau) ,H_I(\tau')\}_0
	+ \int_{\tau_0}^\tau \dd \tau' \int_{\tau_0}^{\tau'} \dd \tau'' \{ \{h_{\vb{k}}^{f,\gamma}(\tau) h_{\vb{k}'}^{f,\gamma'}(\tau) ,H_I(\tau')\}_0,H_I(\tau'')\}_0  \,.
\end{equation}
Using the interaction Hamiltonian from Eq.~(12), we obtain
\begin{align} \label{eq: SIGWs din Cl}
	\nonumber
	&h_{\vb{k}}^\gamma h_{\vb{k}'}^{\gamma'} =  h_{\vb{k}}^{f,\gamma} h_{\vb{k}'}^{f,\gamma'} + \Bigg[ -\dfrac{1}{2} \int \dfrac{\dd^3 \vb{p}}{(2\pi)^{3/2}}  e^{\gamma'}_{ij}(\vb{k}') \vb{p}_{i} \vb{p}_{j} \,  h^{f,\gamma}_{\vb{k}} \int_{\tau_0}^\tau \dd \tau' \, g_{k'}^h(\tau,\tau') \chi^f_{c,\vb{k}'-\vb{p}} \, \chi^f_{c,\vb{p}}  \\\nonumber
	&-\dfrac{1}{2 } \int \dfrac{\dd^3 \vb{p}\, \dd^3 \vb{q} }{(2\pi)^3} \sum_{\gamma''} e^{\gamma'}_{ik}(\vb{k}') \vb{q}_{i} \, e^{\gamma''}_{jk}(\vb{p}) (\vb{k}'-\vb{q})_{j} \, h^{f,\gamma}_{\vb{k}} \int_{\tau_0}^\tau \dd \tau' \, g_{k'}^h(\tau,\tau') h^{f,\gamma''}_{\vb{p}} \, \chi^f_{c,\vb{q}} \, \chi^f_{c,\vb{k}'-\vb{p}-\vb{q}}\\ \nonumber
	& + \dfrac{1}{4} \int  \dfrac{\dd^3 \vb{p} \,  \dd^3 \vb{q}}{(2\pi)^{3}} e^{\gamma'}_{ij}(\vb{k}') \vb{p}_{i} \vb{p}_{j}   e^\gamma_{kl}(\vb{k}) \vb{q}_{k} \vb{q}_{l}  \int_{\tau_0}^\tau \dd \tau' \,   g^h_{k'}(\tau,\tau') \chi^f_{c,\vb{p}}\, \chi^f_{c,\vb{k}'-\vb{p}} \int_{\tau_0}^{\tau'} \dd \tau'' \,   g^h_k(\tau,\tau'') \chi^f_{c,\vb{q}}\, \chi^f_{c,\vb{k}-\vb{q}} \\ 
	& + \int  \dfrac{\dd^3 \vb{p}\, \dd^3 \vb{q}}{(2\pi)^{3}}  e^{\gamma'}_{ij}(\vb{k}') \vb{p}_{i} \vb{p}_{j}   \sum_{\gamma''} e^{\gamma''}_{kl}(\vb{q}) \vb{p}_{k}\vb{p}_l\,
	 h^{f,\gamma}_{\vb{k}}\int_{\tau_0}^\tau \dd \tau' \,  g^h_{k'}(\tau,\tau') \chi^f_{c,\vb{k}'-\vb{p}} \int_{\tau_0}^{\tau'} \dd \tau'' \,  g^\chi_{p}(\tau',\tau'') h^{f,\gamma''}_{\vb{q}} \chi^f_{c,\vb{p}-\vb{q}} +(\vb{k},\gamma \leftrightarrow \vb{k}',\gamma') \Bigg] \,,
\end{align}
where we have used
\begin{align}
	&\{h_{\vb{k}}^{f,\gamma}(\tau), h_{\vb{k}'}^{f,\gamma}(\tau')\}_0 = \dfrac{2}{\hbar} \Im{\tilde{s}_k(\tau) \tilde{s}_k^*(\tau')} \delta_{\gamma,\gamma'} \delta(\vb{k} + \vb{k}') \equiv g_k^h(\tau,\tau') \delta_{\gamma,\gamma'} \delta(\vb{k} + \vb{k}') \,,\\
	& \{\chi^f_{c,\vb{k}}(\tau), \chi^f_{c,\vb{k}'}(\tau')\}_0 = \dfrac{2}{\hbar} \Im{s_{k}(\tau) s_{k}^*(\tau')} \delta(\vb{k} + \vb{k}') \equiv g_k^\chi(\tau,\tau') \delta(\vb{k} + \vb{k}') \,.
\end{align} 
To lighten the notation, every function inside an integrand in \eqref{eq: SIGWs din Cl} is to be evaluated in the corresponding integration time, unless otherwise explicitly stated. Similarly, functions outside integrals are evaluated at $\tau$.

To compute the classical average of $h_{\vb{k}}^\gamma h_{\vb{k}'}^{\gamma'}$ from \eqref{eq: SIGWs din Cl}, we just need to specify the statistics of $\alpha_{\vb{k}}$ and $\alpha_{\vb{k}}^\gamma$. As explained in the main text, these random variables are related to the initial conditions of the fields and their conjugate momenta in the interaction picture, $(\Pi_{\chi_c})_{\vb{k}} = \chi'_{c,\vb{k}}$ and $(\Pi_{h})^\gamma_{\vb{k}} = (a^2M_P^2/4) h'^\gamma_{\vb{k}}$:
\begin{align}\label{ec:relac_alpha_IC}
	\alpha_{\vb{k}} = \dfrac{1}{i \hbar}\Big[ \chi_{c,\vb{k}}(\tau_0) s'^*_{k}(\tau_0) - \Pi_{\chi_c,\vb{k}}(\tau_0) s^*_{k}(\tau_0) \Big]\,,\quad
	\alpha^\gamma_{\vb{k}} = \dfrac{1}{i \hbar}\Big[ \dfrac{a^2(\tau_0) M_P^2 }{4} h^{\gamma}_{\vb{k}}(\tau_0) \tilde{s}'^{*}_{k}(\tau_0) - \Pi^{\gamma}_{h,\vb{k}}(\tau_0) \tilde{s}^*_{k}(\tau_0) \Big] \,,
\end{align}
where we used the normalization condition for the Wronskian of mode functions of canonically normalized fields: 
\begin{equation}
	s_{k} s'^*_{k} - s'_{k} s^*_{k} = i \hbar \quad {\rm and} \quad \dfrac{a^2M_P^2}{4}(\tilde{s}_{k} \tilde{s}'^*_{k} - \tilde s'_{k} \tilde s^*_{k}) = i \hbar \,.
\end{equation}
From Eq.~\eqref{ec:relac_alpha_IC}, and imposing that the averages of products of classical initial fields and momenta coincide with the quantum expectation values of the corresponding products of field and momentum operators evaluated at $\tau=\tau_0$, one can obtain the statistics of $\alpha_{\vb{k}}$ and $\alpha_{\vb{k}}^\gamma$. Let us explain in some detail how this is done for the specific case of the product $\alpha^\gamma_{\vb{k}} \alpha^{\gamma'}_{\vb{k}'}$. Expanding at second order in the interaction Hamiltonian:
\begin{align}
	\nonumber
	&\expval{\alpha^\gamma_{\vb{k}} \alpha^{\gamma'}_{\vb{k}'}}_c^{\rm Cl} = \bra{0} \mathcal{S}[\alpha^\gamma_{\vb{k}} \alpha^{\gamma'}_{\vb{k}'}] \ket{0}-\dfrac{i}{\hbar} \int_{-\infty_-}^{\tau_0} \dd \tau' \bra{0} \mathcal{S}[\alpha^\gamma_{\vb{k}} \alpha^{\gamma'}_{\vb{k}'}] H_I(\tau') \ket{0} +\dfrac{i}{\hbar} \int_{-\infty_+}^{\tau_0} \dd \tau' \bra{0} H_I(\tau') \mathcal{S}[\alpha^\gamma_{\vb{k}} \alpha^{\gamma'}_{\vb{k}'}]  \ket{0} \\ \nonumber
	&- \dfrac{1}{\hbar^2} \int_{-\infty_-}^{\tau_0} \mkern-10mu \dd \tau' \int_{-\infty_-}^{\tau'} \mkern-10mu \dd \tau''  \bra{0}  \mathcal{S}[\alpha^\gamma_{\vb{k}} \alpha^{\gamma'}_{\vb{k}'}] H_I(\tau') H_I(\tau'')  \ket{0} - \dfrac{1}{\hbar^2} \int_{-\infty_+}^{\tau_0} \mkern-10mu \dd \tau' \int_{-\infty_+}^{\tau'} \mkern-10mu \dd \tau''  \bra{0}  H_I(\tau'') H_I(\tau')  \mathcal{S}[\alpha^\gamma_{\vb{k}} \alpha^{\gamma'}_{\vb{k}'}]  \ket{0} \\ 
	&+\dfrac{1}{\hbar^2} \int_{-\infty_+}^{\tau_0} \dd \tau' \int_{-\infty_-}^{\tau_0} \dd \tau'' \bra{0} H_I(\tau') \mathcal{S}[\alpha^\gamma_{\vb{k}} \alpha^{\gamma'}_{\vb{k}'}] H_I(\tau'') \ket{0} + \dots\,,
\end{align}
where the symmetrizer $\mathcal{S}[AB]\equiv(AB+BA)/2$ is included since, upon quantization, we must explicitly symmetrize. Despite the apparent complexity, the evaluation is straightforward because many terms vanish. Using Wick's theorem, any $n$-point correlator of free fields reduces to sums over products of two-point contractions. When $\alpha^\gamma_{\vb{k}}$ appears on the left of a contraction, the fields effectively evaluate to the mode function $\tilde s_k$; when it appears on the right, they evaluate to $\tilde s_k^*$. Given the specific combination defining $\alpha^\gamma_{\vb{k}}$, the latter contributions vanish. Therefore only terms in which $\alpha^\gamma_{\vb{k}}$ appears on the left survive, yielding
\begin{align}
	\expval{\alpha^\gamma_{\vb{k}} \alpha^{\gamma'}_{\vb{k}'}}_c^{\rm Cl} = -\dfrac{i}{\hbar} \int_{-\infty_-}^{\tau_0} \dd \tau' \bra{0} \alpha^\gamma_{\vb{k}} \alpha^{\gamma'}_{\vb{k}'} H_I(\tau') \ket{0}- \dfrac{1}{\hbar^2} \int_{-\infty_-}^{\tau_0} \mkern-10mu \dd \tau' \int_{-\infty_-}^{\tau'} \mkern-10mu \dd \tau''  \bra{0}  \alpha^\gamma_{\vb{k}} \alpha^{\gamma'}_{\vb{k}'} H_I(\tau') H_I(\tau'')  \ket{0}\,.
	\end{align}
The symmetrizer is trivial for connected contributions, since the external operators ($\alpha^\gamma_{\vb{k}} \alpha^{\gamma'}_{\vb{k}'}$) contract with the Hamiltonians rather than with each other. The first term receives contributions only from the quartic Hamiltonian, while the second involves two cubic insertions, consistently with working at order $M_P^{-4}$. One finds
	\begin{align} \label{ec:average_alpha}
	\expval{\alpha^\gamma_{\vb{k}} \alpha^{\gamma'}_{\vb{k}'}}_c^{\rm Cl} &= \delta(\vb{k} + \vb{k'}) \int \dfrac{\dd^3 \vb{p}}{(2\pi)^3} \Bigg[ -\dfrac{i}{2\hbar} e^\gamma_{ik}(\vb{k}) e^{\gamma'}_{kj}(\vb{k}) \vb{p}_i \vb{p}_j \int_{-\infty_-}^{\tau_0} \dd \tau' \tilde{s}^{*2}_k \abs{s_{p}}^2 \nonumber\\
	& \quad\qquad\qquad\qquad\qquad\quad - \dfrac{1}{\hbar^2}  e^\gamma_{ij}(\vb{k})\vb{p}_i \vb{p}_j \, e^{\gamma'}_{kl}(\vb{k})\vb{p}_k\vb{p}_l \int_{-\infty_-}^{\tau_0} \mkern-10mu \dd \tau'  \tilde{s}_k^* s_{p} s_{q} \int_{-\infty_-}^{\tau'} \mkern-10mu \dd \tau'' \tilde{s}_k^* s^*_{p} s^*_{q} \Bigg]\,.
\end{align}
Systematically applying the procedure leading to \eqref{ec:average_alpha}, we obtain the other non-zero classical averages of products of $\alpha_{\vb{k}}$ and $\alpha_{\vb{k}}^\gamma$ needed to ensure the equality of the quantum and classical tensor power spectra at $\order{M_P^{-4}}$ and at $\tau=\tau_0$:
\begin{align}
\expval{\alpha_{\vb{k}} \alpha^{*}_{-\vb{k}'}}_c^{\rm Cl} &= \dfrac{1 }{2} \delta(\vb{k} + \vb{k}') \,,\\
\expval{\alpha^\gamma_{\vb{k}} \alpha^{\gamma',*}_{-\vb{k}'}}_c^{\rm Cl}  &= \delta(\vb{k} + \vb{k}') \Bigg[ \dfrac{\delta_{\gamma,\gamma'} }{2}  +  \int \dfrac{\dd^3 \vb{p}}{(2\pi)^3}  \dfrac{1}{2\hbar^2}  e^\gamma_{ij}(\vb{k})\vb{p}_i \vb{p}_j \, e^{\gamma'}_{kl}(\vb{k})\vb{p}_k\vb{p}_l \int_{-\infty_+}^{\tau_0} \mkern-10mu \dd \tau'  \tilde{s}_k s_{p} s_{q} \int_{-\infty_-}^{\tau_0} \mkern-10mu \dd \tau'' \tilde{s}_k^* s^*_{p} s^*_{q} \Bigg]\,,\\
\expval{\alpha_{\vb{k}}^{\gamma} \alpha_{\vb{p}} \alpha_{\vb{q}}}^{\rm Cl}_c &= -\dfrac{i}{\hbar} \int_{-\infty_-}^{\tau_0} \mkern-10mu \dd \tau' \bra{0} \alpha_{\vb{k}}^{\gamma} \alpha_{\vb{p}} \alpha_{\vb{q}} H_I(\tau') \ket{0} =  \dfrac{i}{\hbar} \dfrac{\delta(\vb{k} + \vb{p} + \vb{q})}{(2\pi)^{3/2}} e^\gamma_{ij}(\vb{k}) \vb{p}_i \vb{p}_j \int_{-\infty_-}^{\tau_0} \mkern-10mu \dd \tau' \tilde{s}_k^ * s_{p}^* s_{q}^*\,.
\end{align}
Using these results in the average of \eqref{eq: SIGWs din Cl}, we obtain the classical tensor power spectrum at one loop (i.e.\ at $\order{M_P^{-4}}$) arising from a stochastic field with a power spectrum and a bispectrum coinciding with the quantum ones at $\tau=\tau_0$. Subtracting this from the quantum result for the dimensionless power spectrum of tensor modes (see \cite{Ballesteros:2024cef}),
\begin{align} \nonumber
	\mathcal{P}_h(\tau,k) & = \frac{k^{3}}{4\pi^2} \int  \frac{\dd^{3} \vb{p}}{(2\pi)^3} \Bigg[ p^2 \sin^2\theta  \Im{\tilde s_{k}^2 \int_{-\infty_-}^\tau \dd \tau' \, \tilde s_{k}^{*2} \abs{s_{p}}^2}\\ \nonumber + & \frac{p^4}{2}   \sin^4\theta \abs{\tilde s_k}^2  \int_{-\infty_+}^\tau \mkern-10mu \dd \tau' \, \tilde s_k s_{p} s_{q} \int_{-\infty_-}^\tau \mkern-10mu \dd \tau'' \, \tilde s^*_k  s_{p}^*  s_{q}^* 
	+ \frac{p^4 }{2}  \sin^4\theta  \Re \bigg\{ -2\tilde s_k^2 \int_{-\infty_-}^\tau \mkern-10mu \dd \tau' \, \tilde s^*_k s_{p} s_{q}  \int_{-\infty_-}^{\tau'} \mkern-10mu \dd \tau'' \, \tilde s_k^*   s_{p}^* s_{q}^* \bigg\}\Bigg]\,, 
\end{align} 
Eq.~(13) is obtained.

\bibliographystyle{apsrev4-2}
\bibliography{biblio}

\begin{thebibliography}{56}%
\makeatletter
\providecommand \@ifxundefined [1]{%
 \@ifx{#1\undefined}
}%
\providecommand \@ifnum [1]{%
 \ifnum #1\expandafter \@firstoftwo
 \else \expandafter \@secondoftwo
 \fi
}%
\providecommand \@ifx [1]{%
 \ifx #1\expandafter \@firstoftwo
 \else \expandafter \@secondoftwo
 \fi
}%
\providecommand \natexlab [1]{#1}%
\providecommand \enquote  [1]{``#1''}%
\providecommand \bibnamefont  [1]{#1}%
\providecommand \bibfnamefont [1]{#1}%
\providecommand \citenamefont [1]{#1}%
\providecommand \href@noop [0]{\@secondoftwo}%
\providecommand \href [0]{\begingroup \@sanitize@url \@href}%
\providecommand \@href[1]{\@@startlink{#1}\@@href}%
\providecommand \@@href[1]{\endgroup#1\@@endlink}%
\providecommand \@sanitize@url [0]{\catcode `\\12\catcode `\$12\catcode
  `\&12\catcode `\#12\catcode `\^12\catcode `\_12\catcode `\%12\relax}%
\providecommand \@@startlink[1]{}%
\providecommand \@@endlink[0]{}%
\providecommand \url  [0]{\begingroup\@sanitize@url \@url }%
\providecommand \@url [1]{\endgroup\@href {#1}{\urlprefix }}%
\providecommand \urlprefix  [0]{URL }%
\providecommand \Eprint [0]{\href }%
\providecommand \doibase [0]{https://doi.org/}%
\providecommand \selectlanguage [0]{\@gobble}%
\providecommand \bibinfo  [0]{\@secondoftwo}%
\providecommand \bibfield  [0]{\@secondoftwo}%
\providecommand \translation [1]{[#1]}%
\providecommand \BibitemOpen [0]{}%
\providecommand \bibitemStop [0]{}%
\providecommand \bibitemNoStop [0]{.\EOS\space}%
\providecommand \EOS [0]{\spacefactor3000\relax}%
\providecommand \BibitemShut  [1]{\csname bibitem#1\endcsname}%
\let\auto@bib@innerbib\@empty
\bibitem [{\citenamefont {Starobinsky}(1979)}]{Starobinsky:1979ty}%
  \BibitemOpen
  \bibfield  {author} {\bibinfo {author} {\bibfnamefont {A.~A.}\ \bibnamefont
  {Starobinsky}},\ }\href@noop {} {\bibfield  {journal} {\bibinfo  {journal}
  {JETP Lett.}\ }\textbf {\bibinfo {volume} {30}},\ \bibinfo {pages} {682}
  (\bibinfo {year} {1979})}\BibitemShut {NoStop}%
\bibitem [{\citenamefont {Guth}(1981)}]{Guth:1980zm}%
  \BibitemOpen
  \bibfield  {author} {\bibinfo {author} {\bibfnamefont {A.~H.}\ \bibnamefont
  {Guth}},\ }\href {https://doi.org/10.1103/PhysRevD.23.347} {\bibfield
  {journal} {\bibinfo  {journal} {Phys. Rev. D}\ }\textbf {\bibinfo {volume}
  {23}},\ \bibinfo {pages} {347} (\bibinfo {year} {1981})}\BibitemShut
  {NoStop}%
\bibitem [{\citenamefont {Linde}(1982)}]{Linde:1981mu}%
  \BibitemOpen
  \bibfield  {author} {\bibinfo {author} {\bibfnamefont {A.~D.}\ \bibnamefont
  {Linde}},\ }\href {https://doi.org/10.1016/0370-2693(82)91219-9} {\bibfield
  {journal} {\bibinfo  {journal} {Phys. Lett. B}\ }\textbf {\bibinfo {volume}
  {108}},\ \bibinfo {pages} {389} (\bibinfo {year} {1982})}\BibitemShut
  {NoStop}%
\bibitem [{\citenamefont {Peiris}\ \emph {et~al.}(2003)\citenamefont {Peiris}
  \emph {et~al.}}]{WMAP:2003syu}%
  \BibitemOpen
  \bibfield  {author} {\bibinfo {author} {\bibfnamefont {H.~V.}\ \bibnamefont
  {Peiris}} \emph {et~al.} (\bibinfo {collaboration} {WMAP}),\ }\href
  {https://doi.org/10.1086/377228} {\bibfield  {journal} {\bibinfo  {journal}
  {Astrophys. J. Suppl.}\ }\textbf {\bibinfo {volume} {148}},\ \bibinfo {pages}
  {213} (\bibinfo {year} {2003})},\ \Eprint
  {https://arxiv.org/abs/astro-ph/0302225} {arXiv:astro-ph/0302225}
  \BibitemShut {NoStop}%
\bibitem [{\citenamefont {Senatore}\ \emph {et~al.}(2010)\citenamefont
  {Senatore}, \citenamefont {Smith},\ and\ \citenamefont
  {Zaldarriaga}}]{Senatore:2009gt}%
  \BibitemOpen
  \bibfield  {author} {\bibinfo {author} {\bibfnamefont {L.}~\bibnamefont
  {Senatore}}, \bibinfo {author} {\bibfnamefont {K.~M.}\ \bibnamefont
  {Smith}},\ and\ \bibinfo {author} {\bibfnamefont {M.}~\bibnamefont
  {Zaldarriaga}},\ }\href {https://doi.org/10.1088/1475-7516/2010/01/028}
  {\bibfield  {journal} {\bibinfo  {journal} {JCAP}\ }\textbf {\bibinfo
  {volume} {01}},\ \bibinfo {pages} {028}},\ \Eprint
  {https://arxiv.org/abs/0905.3746} {arXiv:0905.3746 [astro-ph.CO]}
  \BibitemShut {NoStop}%
\bibitem [{\citenamefont {Akrami}\ \emph
  {et~al.}(2020{\natexlab{a}})\citenamefont {Akrami} \emph
  {et~al.}}]{Planck:2018jri}%
  \BibitemOpen
  \bibfield  {author} {\bibinfo {author} {\bibfnamefont {Y.}~\bibnamefont
  {Akrami}} \emph {et~al.} (\bibinfo {collaboration} {Planck}),\ }\href
  {https://doi.org/10.1051/0004-6361/201833887} {\bibfield  {journal} {\bibinfo
   {journal} {Astron. Astrophys.}\ }\textbf {\bibinfo {volume} {641}},\
  \bibinfo {pages} {A10} (\bibinfo {year} {2020}{\natexlab{a}})},\ \Eprint
  {https://arxiv.org/abs/1807.06211} {arXiv:1807.06211 [astro-ph.CO]}
  \BibitemShut {NoStop}%
\bibitem [{\citenamefont {Akrami}\ \emph
  {et~al.}(2020{\natexlab{b}})\citenamefont {Akrami} \emph
  {et~al.}}]{Planck:2019kim}%
  \BibitemOpen
  \bibfield  {author} {\bibinfo {author} {\bibfnamefont {Y.}~\bibnamefont
  {Akrami}} \emph {et~al.} (\bibinfo {collaboration} {Planck}),\ }\href
  {https://doi.org/10.1051/0004-6361/201935891} {\bibfield  {journal} {\bibinfo
   {journal} {Astron. Astrophys.}\ }\textbf {\bibinfo {volume} {641}},\
  \bibinfo {pages} {A9} (\bibinfo {year} {2020}{\natexlab{b}})},\ \Eprint
  {https://arxiv.org/abs/1905.05697} {arXiv:1905.05697 [astro-ph.CO]}
  \BibitemShut {NoStop}%
\bibitem [{\citenamefont {Cabass}\ \emph {et~al.}(2022)\citenamefont {Cabass},
  \citenamefont {Ivanov}, \citenamefont {Philcox}, \citenamefont
  {Simonovi{\'c}},\ and\ \citenamefont {Zaldarriaga}}]{Cabass:2022wjy}%
  \BibitemOpen
  \bibfield  {author} {\bibinfo {author} {\bibfnamefont {G.}~\bibnamefont
  {Cabass}}, \bibinfo {author} {\bibfnamefont {M.~M.}\ \bibnamefont {Ivanov}},
  \bibinfo {author} {\bibfnamefont {O.~H.~E.}\ \bibnamefont {Philcox}},
  \bibinfo {author} {\bibfnamefont {M.}~\bibnamefont {Simonovi{\'c}}},\ and\
  \bibinfo {author} {\bibfnamefont {M.}~\bibnamefont {Zaldarriaga}},\ }\href
  {https://doi.org/10.1103/PhysRevLett.129.021301} {\bibfield  {journal}
  {\bibinfo  {journal} {Phys. Rev. Lett.}\ }\textbf {\bibinfo {volume} {129}},\
  \bibinfo {pages} {021301} (\bibinfo {year} {2022})},\ \Eprint
  {https://arxiv.org/abs/2201.07238} {arXiv:2201.07238 [astro-ph.CO]}
  \BibitemShut {NoStop}%
\bibitem [{\citenamefont {D'Amico}\ \emph {et~al.}(2025)\citenamefont
  {D'Amico}, \citenamefont {Lewandowski}, \citenamefont {Senatore},\ and\
  \citenamefont {Zhang}}]{DAmico:2022gki}%
  \BibitemOpen
  \bibfield  {author} {\bibinfo {author} {\bibfnamefont {G.}~\bibnamefont
  {D'Amico}}, \bibinfo {author} {\bibfnamefont {M.}~\bibnamefont
  {Lewandowski}}, \bibinfo {author} {\bibfnamefont {L.}~\bibnamefont
  {Senatore}},\ and\ \bibinfo {author} {\bibfnamefont {P.}~\bibnamefont
  {Zhang}},\ }\href {https://doi.org/10.1103/PhysRevD.111.063514} {\bibfield
  {journal} {\bibinfo  {journal} {Phys. Rev. D}\ }\textbf {\bibinfo {volume}
  {111}},\ \bibinfo {pages} {063514} (\bibinfo {year} {2025})},\ \Eprint
  {https://arxiv.org/abs/2201.11518} {arXiv:2201.11518 [astro-ph.CO]}
  \BibitemShut {NoStop}%
\bibitem [{\citenamefont {Ach{\'u}carro}\ \emph {et~al.}(2022)\citenamefont
  {Ach{\'u}carro} \emph {et~al.}}]{Achucarro:2022qrl}%
  \BibitemOpen
  \bibfield  {author} {\bibinfo {author} {\bibfnamefont {A.}~\bibnamefont
  {Ach{\'u}carro}} \emph {et~al.},\ }\href@noop {} {\  (\bibinfo {year}
  {2022})},\ \Eprint {https://arxiv.org/abs/2203.08128} {arXiv:2203.08128
  [astro-ph.CO]} \BibitemShut {NoStop}%
\bibitem [{\citenamefont {Allys}\ \emph {et~al.}(2023)\citenamefont {Allys}
  \emph {et~al.}}]{LiteBIRD:2022cnt}%
  \BibitemOpen
  \bibfield  {author} {\bibinfo {author} {\bibfnamefont {E.}~\bibnamefont
  {Allys}} \emph {et~al.} (\bibinfo {collaboration} {LiteBIRD}),\ }\href
  {https://doi.org/10.1093/ptep/ptac150} {\bibfield  {journal} {\bibinfo
  {journal} {PTEP}\ }\textbf {\bibinfo {volume} {2023}},\ \bibinfo {pages}
  {042F01} (\bibinfo {year} {2023})},\ \Eprint
  {https://arxiv.org/abs/2202.02773} {arXiv:2202.02773 [astro-ph.IM]}
  \BibitemShut {NoStop}%
\bibitem [{\citenamefont {Liddle}\ and\ \citenamefont
  {Leach}(2003)}]{Liddle:2003as}%
  \BibitemOpen
  \bibfield  {author} {\bibinfo {author} {\bibfnamefont {A.~R.}\ \bibnamefont
  {Liddle}}\ and\ \bibinfo {author} {\bibfnamefont {S.~M.}\ \bibnamefont
  {Leach}},\ }\href {https://doi.org/10.1103/PhysRevD.68.103503} {\bibfield
  {journal} {\bibinfo  {journal} {Phys. Rev. D}\ }\textbf {\bibinfo {volume}
  {68}},\ \bibinfo {pages} {103503} (\bibinfo {year} {2003})},\ \Eprint
  {https://arxiv.org/abs/astro-ph/0305263} {arXiv:astro-ph/0305263}
  \BibitemShut {NoStop}%
\bibitem [{\citenamefont {Gow}\ \emph {et~al.}(2021)\citenamefont {Gow},
  \citenamefont {Byrnes}, \citenamefont {Cole},\ and\ \citenamefont
  {Young}}]{Gow:2020bzo}%
  \BibitemOpen
  \bibfield  {author} {\bibinfo {author} {\bibfnamefont {A.~D.}\ \bibnamefont
  {Gow}}, \bibinfo {author} {\bibfnamefont {C.~T.}\ \bibnamefont {Byrnes}},
  \bibinfo {author} {\bibfnamefont {P.~S.}\ \bibnamefont {Cole}},\ and\
  \bibinfo {author} {\bibfnamefont {S.}~\bibnamefont {Young}},\ }\href
  {https://doi.org/10.1088/1475-7516/2021/02/002} {\bibfield  {journal}
  {\bibinfo  {journal} {JCAP}\ }\textbf {\bibinfo {volume} {02}},\ \bibinfo
  {pages} {002}},\ \Eprint {https://arxiv.org/abs/2008.03289} {arXiv:2008.03289
  [astro-ph.CO]} \BibitemShut {NoStop}%
\bibitem [{\citenamefont {Saito}\ and\ \citenamefont
  {Yokoyama}(2009)}]{Saito:2008jc}%
  \BibitemOpen
  \bibfield  {author} {\bibinfo {author} {\bibfnamefont {R.}~\bibnamefont
  {Saito}}\ and\ \bibinfo {author} {\bibfnamefont {J.}~\bibnamefont
  {Yokoyama}},\ }\href {https://doi.org/10.1103/PhysRevLett.102.161101}
  {\bibfield  {journal} {\bibinfo  {journal} {Phys. Rev. Lett.}\ }\textbf
  {\bibinfo {volume} {102}},\ \bibinfo {pages} {161101} (\bibinfo {year}
  {2009})},\ \bibinfo {note} {[Erratum: Phys.Rev.Lett. 107, 069901 (2011)]},\
  \Eprint {https://arxiv.org/abs/0812.4339} {arXiv:0812.4339 [astro-ph]}
  \BibitemShut {NoStop}%
\bibitem [{\citenamefont {Bartolo}\ \emph {et~al.}(2019)\citenamefont
  {Bartolo}, \citenamefont {De~Luca}, \citenamefont {Franciolini},
  \citenamefont {Lewis}, \citenamefont {Peloso},\ and\ \citenamefont
  {Riotto}}]{Bartolo:2018evs}%
  \BibitemOpen
  \bibfield  {author} {\bibinfo {author} {\bibfnamefont {N.}~\bibnamefont
  {Bartolo}}, \bibinfo {author} {\bibfnamefont {V.}~\bibnamefont {De~Luca}},
  \bibinfo {author} {\bibfnamefont {G.}~\bibnamefont {Franciolini}}, \bibinfo
  {author} {\bibfnamefont {A.}~\bibnamefont {Lewis}}, \bibinfo {author}
  {\bibfnamefont {M.}~\bibnamefont {Peloso}},\ and\ \bibinfo {author}
  {\bibfnamefont {A.}~\bibnamefont {Riotto}},\ }\href
  {https://doi.org/10.1103/PhysRevLett.122.211301} {\bibfield  {journal}
  {\bibinfo  {journal} {Phys. Rev. Lett.}\ }\textbf {\bibinfo {volume} {122}},\
  \bibinfo {pages} {211301} (\bibinfo {year} {2019})},\ \Eprint
  {https://arxiv.org/abs/1810.12218} {arXiv:1810.12218 [astro-ph.CO]}
  \BibitemShut {NoStop}%
\bibitem [{\citenamefont {Caravano}\ \emph {et~al.}(2021)\citenamefont
  {Caravano}, \citenamefont {Komatsu}, \citenamefont {Lozanov},\ and\
  \citenamefont {Weller}}]{Caravano:2021pgc}%
  \BibitemOpen
  \bibfield  {author} {\bibinfo {author} {\bibfnamefont {A.}~\bibnamefont
  {Caravano}}, \bibinfo {author} {\bibfnamefont {E.}~\bibnamefont {Komatsu}},
  \bibinfo {author} {\bibfnamefont {K.~D.}\ \bibnamefont {Lozanov}},\ and\
  \bibinfo {author} {\bibfnamefont {J.}~\bibnamefont {Weller}},\ }\href
  {https://doi.org/10.1088/1475-7516/2021/12/010} {\bibfield  {journal}
  {\bibinfo  {journal} {JCAP}\ }\textbf {\bibinfo {volume} {12}}\bibfield
  {number} {\bibinfo  {number} { (12)},\ \bibinfo {pages} {010}},\ }\Eprint
  {https://arxiv.org/abs/2102.06378} {arXiv:2102.06378 [astro-ph.CO]}
  \BibitemShut {NoStop}%
\bibitem [{\citenamefont {Caravano}\ \emph {et~al.}(2024)\citenamefont
  {Caravano}, \citenamefont {Inomata},\ and\ \citenamefont
  {Renaux-Petel}}]{Caravano:2024tlp}%
  \BibitemOpen
  \bibfield  {author} {\bibinfo {author} {\bibfnamefont {A.}~\bibnamefont
  {Caravano}}, \bibinfo {author} {\bibfnamefont {K.}~\bibnamefont {Inomata}},\
  and\ \bibinfo {author} {\bibfnamefont {S.}~\bibnamefont {Renaux-Petel}},\
  }\href {https://doi.org/10.1103/PhysRevLett.133.151001} {\bibfield  {journal}
  {\bibinfo  {journal} {Phys. Rev. Lett.}\ }\textbf {\bibinfo {volume} {133}},\
  \bibinfo {pages} {151001} (\bibinfo {year} {2024})},\ \Eprint
  {https://arxiv.org/abs/2403.12811} {arXiv:2403.12811 [astro-ph.CO]}
  \BibitemShut {NoStop}%
\bibitem [{\citenamefont {Caravano}\ \emph {et~al.}(2025)\citenamefont
  {Caravano}, \citenamefont {Franciolini},\ and\ \citenamefont
  {Renaux-Petel}}]{Caravano:2024moy}%
  \BibitemOpen
  \bibfield  {author} {\bibinfo {author} {\bibfnamefont {A.}~\bibnamefont
  {Caravano}}, \bibinfo {author} {\bibfnamefont {G.}~\bibnamefont
  {Franciolini}},\ and\ \bibinfo {author} {\bibfnamefont {S.}~\bibnamefont
  {Renaux-Petel}},\ }\href {https://doi.org/10.1103/PhysRevD.111.063518}
  {\bibfield  {journal} {\bibinfo  {journal} {Phys. Rev. D}\ }\textbf {\bibinfo
  {volume} {111}},\ \bibinfo {pages} {063518} (\bibinfo {year} {2025})},\
  \Eprint {https://arxiv.org/abs/2410.23942} {arXiv:2410.23942 [astro-ph.CO]}
  \BibitemShut {NoStop}%
\bibitem [{\citenamefont {Caravano}\ \emph {et~al.}(2026)\citenamefont
  {Caravano}, \citenamefont {Franciolini},\ and\ \citenamefont
  {Renaux-Petel}}]{Caravano:2026hca}%
  \BibitemOpen
  \bibfield  {author} {\bibinfo {author} {\bibfnamefont {A.}~\bibnamefont
  {Caravano}}, \bibinfo {author} {\bibfnamefont {G.}~\bibnamefont
  {Franciolini}},\ and\ \bibinfo {author} {\bibfnamefont {S.}~\bibnamefont
  {Renaux-Petel}},\ }\href@noop {} {\  (\bibinfo {year} {2026})},\ \Eprint
  {https://arxiv.org/abs/2604.03628} {arXiv:2604.03628 [astro-ph.CO]}
  \BibitemShut {NoStop}%
\bibitem [{\citenamefont {Maldacena}(2003)}]{Maldacena:2002vr}%
  \BibitemOpen
  \bibfield  {author} {\bibinfo {author} {\bibfnamefont {J.~M.}\ \bibnamefont
  {Maldacena}},\ }\href {https://doi.org/10.1088/1126-6708/2003/05/013}
  {\bibfield  {journal} {\bibinfo  {journal} {JHEP}\ }\textbf {\bibinfo
  {volume} {05}},\ \bibinfo {pages} {013}},\ \Eprint
  {https://arxiv.org/abs/astro-ph/0210603} {arXiv:astro-ph/0210603}
  \BibitemShut {NoStop}%
\bibitem [{\citenamefont {Weinberg}(2005)}]{Weinberg:2005vy}%
  \BibitemOpen
  \bibfield  {author} {\bibinfo {author} {\bibfnamefont {S.}~\bibnamefont
  {Weinberg}},\ }\href {https://doi.org/10.1103/PhysRevD.72.043514} {\bibfield
  {journal} {\bibinfo  {journal} {Phys. Rev. D}\ }\textbf {\bibinfo {volume}
  {72}},\ \bibinfo {pages} {043514} (\bibinfo {year} {2005})},\ \Eprint
  {https://arxiv.org/abs/hep-th/0506236} {arXiv:hep-th/0506236} \BibitemShut
  {NoStop}%
\bibitem [{\citenamefont {Senatore}\ and\ \citenamefont
  {Zaldarriaga}(2010)}]{Senatore:2009cf}%
  \BibitemOpen
  \bibfield  {author} {\bibinfo {author} {\bibfnamefont {L.}~\bibnamefont
  {Senatore}}\ and\ \bibinfo {author} {\bibfnamefont {M.}~\bibnamefont
  {Zaldarriaga}},\ }\href {https://doi.org/10.1007/JHEP12(2010)008} {\bibfield
  {journal} {\bibinfo  {journal} {JHEP}\ }\textbf {\bibinfo {volume} {12}},\
  \bibinfo {pages} {008}},\ \Eprint {https://arxiv.org/abs/0912.2734}
  {arXiv:0912.2734 [hep-th]} \BibitemShut {NoStop}%
\bibitem [{\citenamefont {Chen}\ \emph {et~al.}(2017)\citenamefont {Chen},
  \citenamefont {Wang},\ and\ \citenamefont {Xianyu}}]{Chen:2017ryl}%
  \BibitemOpen
  \bibfield  {author} {\bibinfo {author} {\bibfnamefont {X.}~\bibnamefont
  {Chen}}, \bibinfo {author} {\bibfnamefont {Y.}~\bibnamefont {Wang}},\ and\
  \bibinfo {author} {\bibfnamefont {Z.-Z.}\ \bibnamefont {Xianyu}},\ }\href
  {https://doi.org/10.1088/1475-7516/2017/12/006} {\bibfield  {journal}
  {\bibinfo  {journal} {JCAP}\ }\textbf {\bibinfo {volume} {12}},\ \bibinfo
  {pages} {006}},\ \Eprint {https://arxiv.org/abs/1703.10166} {arXiv:1703.10166
  [hep-th]} \BibitemShut {NoStop}%
\bibitem [{\citenamefont {Arkani-Hamed}\ \emph {et~al.}(2020)\citenamefont
  {Arkani-Hamed}, \citenamefont {Baumann}, \citenamefont {Lee},\ and\
  \citenamefont {Pimentel}}]{Arkani-Hamed:2018kmz}%
  \BibitemOpen
  \bibfield  {author} {\bibinfo {author} {\bibfnamefont {N.}~\bibnamefont
  {Arkani-Hamed}}, \bibinfo {author} {\bibfnamefont {D.}~\bibnamefont
  {Baumann}}, \bibinfo {author} {\bibfnamefont {H.}~\bibnamefont {Lee}},\ and\
  \bibinfo {author} {\bibfnamefont {G.~L.}\ \bibnamefont {Pimentel}},\ }\href
  {https://doi.org/10.1007/JHEP04(2020)105} {\bibfield  {journal} {\bibinfo
  {journal} {JHEP}\ }\textbf {\bibinfo {volume} {04}},\ \bibinfo {pages}
  {105}},\ \Eprint {https://arxiv.org/abs/1811.00024} {arXiv:1811.00024
  [hep-th]} \BibitemShut {NoStop}%
\bibitem [{\citenamefont {Baumann}\ \emph {et~al.}(2024)\citenamefont
  {Baumann}, \citenamefont {Green}, \citenamefont {Joyce}, \citenamefont
  {Pajer}, \citenamefont {Pimentel}, \citenamefont {Sleight},\ and\
  \citenamefont {Taronna}}]{Baumann:2022jpr}%
  \BibitemOpen
  \bibfield  {author} {\bibinfo {author} {\bibfnamefont {D.}~\bibnamefont
  {Baumann}}, \bibinfo {author} {\bibfnamefont {D.}~\bibnamefont {Green}},
  \bibinfo {author} {\bibfnamefont {A.}~\bibnamefont {Joyce}}, \bibinfo
  {author} {\bibfnamefont {E.}~\bibnamefont {Pajer}}, \bibinfo {author}
  {\bibfnamefont {G.~L.}\ \bibnamefont {Pimentel}}, \bibinfo {author}
  {\bibfnamefont {C.}~\bibnamefont {Sleight}},\ and\ \bibinfo {author}
  {\bibfnamefont {M.}~\bibnamefont {Taronna}},\ }\href
  {https://doi.org/10.21468/SciPostPhysCommRep.1} {\bibfield  {journal}
  {\bibinfo  {journal} {SciPost Phys. Comm. Rep.}\ }\textbf {\bibinfo {volume}
  {2024}},\ \bibinfo {pages} {1} (\bibinfo {year} {2024})},\ \Eprint
  {https://arxiv.org/abs/2203.08121} {arXiv:2203.08121 [hep-th]} \BibitemShut
  {NoStop}%
\bibitem [{\citenamefont {Kiefer}\ \emph {et~al.}(1998)\citenamefont {Kiefer},
  \citenamefont {Polarski},\ and\ \citenamefont {Starobinsky}}]{Kiefer:1998qe}%
  \BibitemOpen
  \bibfield  {author} {\bibinfo {author} {\bibfnamefont {C.}~\bibnamefont
  {Kiefer}}, \bibinfo {author} {\bibfnamefont {D.}~\bibnamefont {Polarski}},\
  and\ \bibinfo {author} {\bibfnamefont {A.~A.}\ \bibnamefont {Starobinsky}},\
  }\href {https://doi.org/10.1142/S0218271898000292} {\bibfield  {journal}
  {\bibinfo  {journal} {Int. J. Mod. Phys. D}\ }\textbf {\bibinfo {volume}
  {7}},\ \bibinfo {pages} {455} (\bibinfo {year} {1998})},\ \Eprint
  {https://arxiv.org/abs/gr-qc/9802003} {arXiv:gr-qc/9802003} \BibitemShut
  {NoStop}%
\bibitem [{\citenamefont {Maldacena}(2016)}]{Maldacena:2015bha}%
  \BibitemOpen
  \bibfield  {author} {\bibinfo {author} {\bibfnamefont {J.}~\bibnamefont
  {Maldacena}},\ }\href {https://doi.org/10.1002/prop.201500097} {\bibfield
  {journal} {\bibinfo  {journal} {Fortsch. Phys.}\ }\textbf {\bibinfo {volume}
  {64}},\ \bibinfo {pages} {10} (\bibinfo {year} {2016})},\ \Eprint
  {https://arxiv.org/abs/1508.01082} {arXiv:1508.01082 [hep-th]} \BibitemShut
  {NoStop}%
\bibitem [{\citenamefont {Campo}\ and\ \citenamefont
  {Parentani}(2006)}]{Campo:2005sv}%
  \BibitemOpen
  \bibfield  {author} {\bibinfo {author} {\bibfnamefont {D.}~\bibnamefont
  {Campo}}\ and\ \bibinfo {author} {\bibfnamefont {R.}~\bibnamefont
  {Parentani}},\ }\href {https://doi.org/10.1103/PhysRevD.74.025001} {\bibfield
   {journal} {\bibinfo  {journal} {Phys. Rev. D}\ }\textbf {\bibinfo {volume}
  {74}},\ \bibinfo {pages} {025001} (\bibinfo {year} {2006})},\ \Eprint
  {https://arxiv.org/abs/astro-ph/0505376} {arXiv:astro-ph/0505376}
  \BibitemShut {NoStop}%
\bibitem [{\citenamefont {Martin}\ and\ \citenamefont
  {Vennin}(2017)}]{Martin:2017zxs}%
  \BibitemOpen
  \bibfield  {author} {\bibinfo {author} {\bibfnamefont {J.}~\bibnamefont
  {Martin}}\ and\ \bibinfo {author} {\bibfnamefont {V.}~\bibnamefont
  {Vennin}},\ }\href {https://doi.org/10.1103/PhysRevD.96.063501} {\bibfield
  {journal} {\bibinfo  {journal} {Phys. Rev. D}\ }\textbf {\bibinfo {volume}
  {96}},\ \bibinfo {pages} {063501} (\bibinfo {year} {2017})},\ \Eprint
  {https://arxiv.org/abs/1706.05001} {arXiv:1706.05001 [astro-ph.CO]}
  \BibitemShut {NoStop}%
\bibitem [{\citenamefont {Launay}\ \emph {et~al.}(2025)\citenamefont {Launay},
  \citenamefont {Rigopoulos},\ and\ \citenamefont {Shellard}}]{Launay:2024trh}%
  \BibitemOpen
  \bibfield  {author} {\bibinfo {author} {\bibfnamefont {Y.~L.}\ \bibnamefont
  {Launay}}, \bibinfo {author} {\bibfnamefont {G.~I.}\ \bibnamefont
  {Rigopoulos}},\ and\ \bibinfo {author} {\bibfnamefont {E.~P.~S.}\
  \bibnamefont {Shellard}},\ }\href
  {https://doi.org/10.1088/1475-7516/2025/05/071} {\bibfield  {journal}
  {\bibinfo  {journal} {JCAP}\ }\textbf {\bibinfo {volume} {05}},\ \bibinfo
  {pages} {071}},\ \Eprint {https://arxiv.org/abs/2412.16143} {arXiv:2412.16143
  [gr-qc]} \BibitemShut {NoStop}%
\bibitem [{\citenamefont {Ireland}\ and\ \citenamefont
  {Vennin}(2026)}]{Ireland:2026txt}%
  \BibitemOpen
  \bibfield  {author} {\bibinfo {author} {\bibfnamefont {A.}~\bibnamefont
  {Ireland}}\ and\ \bibinfo {author} {\bibfnamefont {V.}~\bibnamefont
  {Vennin}},\ }\href@noop {} {\  (\bibinfo {year} {2026})},\ \Eprint
  {https://arxiv.org/abs/2601.22219} {arXiv:2601.22219 [gr-qc]} \BibitemShut
  {NoStop}%
\bibitem [{\citenamefont {Green}\ and\ \citenamefont
  {Porto}(2020)}]{Green:2020whw}%
  \BibitemOpen
  \bibfield  {author} {\bibinfo {author} {\bibfnamefont {D.}~\bibnamefont
  {Green}}\ and\ \bibinfo {author} {\bibfnamefont {R.~A.}\ \bibnamefont
  {Porto}},\ }\href {https://doi.org/10.1103/PhysRevLett.124.251302} {\bibfield
   {journal} {\bibinfo  {journal} {Phys. Rev. Lett.}\ }\textbf {\bibinfo
  {volume} {124}},\ \bibinfo {pages} {251302} (\bibinfo {year} {2020})},\
  \Eprint {https://arxiv.org/abs/2001.09149} {arXiv:2001.09149 [hep-th]}
  \BibitemShut {NoStop}%
\bibitem [{\citenamefont {Ghosh}\ \emph {et~al.}(2023)\citenamefont {Ghosh},
  \citenamefont {Singh},\ and\ \citenamefont {Ullah}}]{Ghosh:2022cny}%
  \BibitemOpen
  \bibfield  {author} {\bibinfo {author} {\bibfnamefont {D.}~\bibnamefont
  {Ghosh}}, \bibinfo {author} {\bibfnamefont {A.~H.}\ \bibnamefont {Singh}},\
  and\ \bibinfo {author} {\bibfnamefont {F.}~\bibnamefont {Ullah}},\ }\href
  {https://doi.org/10.1088/1475-7516/2023/04/007} {\bibfield  {journal}
  {\bibinfo  {journal} {JCAP}\ }\textbf {\bibinfo {volume} {04}},\ \bibinfo
  {pages} {007}},\ \Eprint {https://arxiv.org/abs/2207.06430} {arXiv:2207.06430
  [hep-th]} \BibitemShut {NoStop}%
\bibitem [{\citenamefont {Creminelli}\ and\ \citenamefont
  {Zaldarriaga}(2004)}]{Creminelli:2004yq}%
  \BibitemOpen
  \bibfield  {author} {\bibinfo {author} {\bibfnamefont {P.}~\bibnamefont
  {Creminelli}}\ and\ \bibinfo {author} {\bibfnamefont {M.}~\bibnamefont
  {Zaldarriaga}},\ }\href {https://doi.org/10.1088/1475-7516/2004/10/006}
  {\bibfield  {journal} {\bibinfo  {journal} {JCAP}\ }\textbf {\bibinfo
  {volume} {10}},\ \bibinfo {pages} {006}},\ \Eprint
  {https://arxiv.org/abs/astro-ph/0407059} {arXiv:astro-ph/0407059}
  \BibitemShut {NoStop}%
\bibitem [{Note1()}]{Note1}%
  \BibitemOpen
  \bibinfo {note} {It was argued in \cite {Green:2020whw} (independently of the
  treatment of the $\tau _0\rightarrow - \infty $ limit) that scenarios with
  dissipative interactions {\cite
  {Berera:1995ie,Berera:1995wh,LopezNacir:2011kk}} or with excited states \cite
  {Holman:2007na,Flauger:2013hra} can be considered classical and lead to
  enhanced folded configurations. In our present work we are not concerned with
  these situations, {as we focus on single-field inflation.} Besides, if
  classical dynamics were to be associated to abundant and continuous particle
  production, it would appear reasonable to set a finite time from which
  sufficient particles are produced. We also note that the divergences found in
  de Sitter models with $\alpha $-vacua \cite {Ghosh:2022cny,Shukla:2016bnu}
  have a closely related origin: the corresponding mode functions contain both
  positive- and negative-frequency phases.}\BibitemShut {Stop}%
\bibitem [{\citenamefont {Ballesteros}\ and\ \citenamefont
  {Gamb{\'\i}n~Egea}(2024)}]{Ballesteros:2024zdp}%
  \BibitemOpen
  \bibfield  {author} {\bibinfo {author} {\bibfnamefont {G.}~\bibnamefont
  {Ballesteros}}\ and\ \bibinfo {author} {\bibfnamefont {J.}~\bibnamefont
  {Gamb{\'\i}n~Egea}},\ }\href {https://doi.org/10.1088/1475-7516/2024/07/052}
  {\bibfield  {journal} {\bibinfo  {journal} {JCAP}\ }\textbf {\bibinfo
  {volume} {07}},\ \bibinfo {pages} {052}},\ \Eprint
  {https://arxiv.org/abs/2404.07196} {arXiv:2404.07196 [astro-ph.CO]}
  \BibitemShut {NoStop}%
\bibitem [{\citenamefont {Ballesteros}\ \emph {et~al.}(2025)\citenamefont
  {Ballesteros}, \citenamefont {Gamb{\'\i}n~Egea},\ and\ \citenamefont
  {Riccardi}}]{Ballesteros:2024cef}%
  \BibitemOpen
  \bibfield  {author} {\bibinfo {author} {\bibfnamefont {G.}~\bibnamefont
  {Ballesteros}}, \bibinfo {author} {\bibfnamefont {J.}~\bibnamefont
  {Gamb{\'\i}n~Egea}},\ and\ \bibinfo {author} {\bibfnamefont {F.}~\bibnamefont
  {Riccardi}},\ }\href {https://doi.org/10.1007/JHEP06(2025)098} {\bibfield
  {journal} {\bibinfo  {journal} {JHEP}\ }\textbf {\bibinfo {volume} {06}},\
  \bibinfo {pages} {098}},\ \Eprint {https://arxiv.org/abs/2411.19674}
  {arXiv:2411.19674 [hep-th]} \BibitemShut {NoStop}%
\bibitem [{\citenamefont {Jamieson}\ \emph {et~al.}(2025)\citenamefont
  {Jamieson}, \citenamefont {Caravano},\ and\ \citenamefont
  {Komatsu}}]{Jamieson:2025ngu}%
  \BibitemOpen
  \bibfield  {author} {\bibinfo {author} {\bibfnamefont {D.}~\bibnamefont
  {Jamieson}}, \bibinfo {author} {\bibfnamefont {A.}~\bibnamefont {Caravano}},\
  and\ \bibinfo {author} {\bibfnamefont {E.}~\bibnamefont {Komatsu}},\ }\href
  {https://doi.org/10.1103/3zml-71jd} {\bibfield  {journal} {\bibinfo
  {journal} {Phys. Rev. D}\ }\textbf {\bibinfo {volume} {112}},\ \bibinfo
  {pages} {103531} (\bibinfo {year} {2025})},\ \Eprint
  {https://arxiv.org/abs/2507.22285} {arXiv:2507.22285 [astro-ph.CO]}
  \BibitemShut {NoStop}%
\bibitem [{\citenamefont {Ballesteros}\ and\ \citenamefont
  {Taoso}(2018)}]{Ballesteros:2017fsr}%
  \BibitemOpen
  \bibfield  {author} {\bibinfo {author} {\bibfnamefont {G.}~\bibnamefont
  {Ballesteros}}\ and\ \bibinfo {author} {\bibfnamefont {M.}~\bibnamefont
  {Taoso}},\ }\href {https://doi.org/10.1103/PhysRevD.97.023501} {\bibfield
  {journal} {\bibinfo  {journal} {Phys. Rev. D}\ }\textbf {\bibinfo {volume}
  {97}},\ \bibinfo {pages} {023501} (\bibinfo {year} {2018})},\ \Eprint
  {https://arxiv.org/abs/1709.05565} {arXiv:1709.05565 [hep-ph]} \BibitemShut
  {NoStop}%
\bibitem [{Note2()}]{Note2}%
  \BibitemOpen
  \bibinfo {note} {The situation can differ for initially excited
  (non-Bunch-Davies) states, for which the real part may be enhanced while the
  imaginary part is fixed by the commutator, equivalently by the Green’s
  function, and is therefore insensitive to the particular
  excitation.}\BibitemShut {Stop}%
\bibitem [{Note3()}]{Note3}%
  \BibitemOpen
  \bibinfo {note} {The generalization to fields follows from taking into
  account that, in Fourier space, fields correspond to a set of coupled
  particles labelled with the continuous index $\vb {k}$ instead of the
  discrete one $a$.}\BibitemShut {Stop}%
\bibitem [{Note4()}]{Note4}%
  \BibitemOpen
  \bibinfo {note} {If $\protect \mathcal {O}$ depended explicitly on time, an
  extra term $\partial \protect \mathcal {O}/\partial t$ would be added on the
  right-hand side.}\BibitemShut {Stop}%
\bibitem [{Note5()}]{Note5}%
  \BibitemOpen
  \bibinfo {note} {Formally, such operator can be related to the time evolution
  operators in the full and free theories as $U(t,t_0) U_f^{-1}(t,t_0) \equiv
  F(t,t_0)$, as it follows from \protect \eqref {ec:evol_of} and \protect
  \eqref {ec:evol_o}.}\BibitemShut {Stop}%
\bibitem [{\citenamefont {Nikolic}(1998)}]{Nikolic:1998tp}%
  \BibitemOpen
  \bibfield  {author} {\bibinfo {author} {\bibfnamefont {H.}~\bibnamefont
  {Nikolic}},\ }\href@noop {} {\  (\bibinfo {year} {1998})},\ \Eprint
  {https://arxiv.org/abs/quant-ph/9812015} {arXiv:quant-ph/9812015}
  \BibitemShut {NoStop}%
\bibitem [{\citenamefont {de~Alwis}(2015)}]{deAlwis:2015ioa}%
  \BibitemOpen
  \bibfield  {author} {\bibinfo {author} {\bibfnamefont {S.~P.}\ \bibnamefont
  {de~Alwis}},\ }\href@noop {} {\  (\bibinfo {year} {2015})},\ \Eprint
  {https://arxiv.org/abs/1504.05211} {arXiv:1504.05211 [hep-th]} \BibitemShut
  {NoStop}%
\bibitem [{Note6()}]{Note6}%
  \BibitemOpen
  \bibinfo {note} {This is the case for a free Hamiltonian $H_f=f_p(t)
  p_f^2+f_q(t) q_f^2$ (we write it for a single particle for simplicity).
  Indeed, in this case the Euler-Lagrange equations are linear, and therefore a
  general solution can be written as a linear combination of two independent
  solutions; in particular, a complex solution and its complex conjugate. Then,
  we can write the general solution for $q_f$ analogously to Eq.~(3), rewrite
  the coefficients of the linear combination of solutions (in Eq.~(3), $\alpha
  $ and $\alpha ^*$) in terms of $q^0$ and $p^0$, and check explicitly that
  $\{q_f(t),p_f(t)\}_0 =1$.}\BibitemShut {Stop}%
\bibitem [{Note7()}]{Note7}%
  \BibitemOpen
  \bibinfo {note} {By assuming time evolution lies on the operators rather than
  the states, we are working in the Heisenberg picture, as usually done in
  cosmology.}\BibitemShut {Stop}%
\bibitem [{Note8()}]{Note8}%
  \BibitemOpen
  \bibinfo {note} {In this respect, we note that the formalism discussed in
  \cite {Musso:2006pt} --written in terms of the Lagrangian, but equivalent to
  the Hamiltonian treatment adopted here, and extended to include the
  $i\epsilon $ prescription in \cite {Adshead:2009cb, Senatore:2009cf}--
  describes interacting quantum fields in a manner formally analogous to
  classical field theory, which is possible since the equations of motion
  formally coincide. However, the formalism in \cite {Musso:2006pt} is purely
  quantum, as it involves (non-commuting) quantum operators rather than
  classical (commuting) functions.}\BibitemShut {Stop}%
\bibitem [{\citenamefont {Berera}(1995)}]{Berera:1995ie}%
  \BibitemOpen
  \bibfield  {author} {\bibinfo {author} {\bibfnamefont {A.}~\bibnamefont
  {Berera}},\ }\href {https://doi.org/10.1103/PhysRevLett.75.3218} {\bibfield
  {journal} {\bibinfo  {journal} {Phys. Rev. Lett.}\ }\textbf {\bibinfo
  {volume} {75}},\ \bibinfo {pages} {3218} (\bibinfo {year} {1995})},\ \Eprint
  {https://arxiv.org/abs/astro-ph/9509049} {arXiv:astro-ph/9509049}
  \BibitemShut {NoStop}%
\bibitem [{\citenamefont {Berera}\ and\ \citenamefont
  {Fang}(1995)}]{Berera:1995wh}%
  \BibitemOpen
  \bibfield  {author} {\bibinfo {author} {\bibfnamefont {A.}~\bibnamefont
  {Berera}}\ and\ \bibinfo {author} {\bibfnamefont {L.-Z.}\ \bibnamefont
  {Fang}},\ }\href {https://doi.org/10.1103/PhysRevLett.74.1912} {\bibfield
  {journal} {\bibinfo  {journal} {Phys. Rev. Lett.}\ }\textbf {\bibinfo
  {volume} {74}},\ \bibinfo {pages} {1912} (\bibinfo {year} {1995})},\ \Eprint
  {https://arxiv.org/abs/astro-ph/9501024} {arXiv:astro-ph/9501024}
  \BibitemShut {NoStop}%
\bibitem [{\citenamefont {Lopez~Nacir}\ \emph {et~al.}(2012)\citenamefont
  {Lopez~Nacir}, \citenamefont {Porto}, \citenamefont {Senatore},\ and\
  \citenamefont {Zaldarriaga}}]{LopezNacir:2011kk}%
  \BibitemOpen
  \bibfield  {author} {\bibinfo {author} {\bibfnamefont {D.}~\bibnamefont
  {Lopez~Nacir}}, \bibinfo {author} {\bibfnamefont {R.~A.}\ \bibnamefont
  {Porto}}, \bibinfo {author} {\bibfnamefont {L.}~\bibnamefont {Senatore}},\
  and\ \bibinfo {author} {\bibfnamefont {M.}~\bibnamefont {Zaldarriaga}},\
  }\href {https://doi.org/10.1007/JHEP01(2012)075} {\bibfield  {journal}
  {\bibinfo  {journal} {JHEP}\ }\textbf {\bibinfo {volume} {01}},\ \bibinfo
  {pages} {075}},\ \Eprint {https://arxiv.org/abs/1109.4192} {arXiv:1109.4192
  [hep-th]} \BibitemShut {NoStop}%
\bibitem [{\citenamefont {Holman}\ and\ \citenamefont
  {Tolley}(2008)}]{Holman:2007na}%
  \BibitemOpen
  \bibfield  {author} {\bibinfo {author} {\bibfnamefont {R.}~\bibnamefont
  {Holman}}\ and\ \bibinfo {author} {\bibfnamefont {A.~J.}\ \bibnamefont
  {Tolley}},\ }\href {https://doi.org/10.1088/1475-7516/2008/05/001} {\bibfield
   {journal} {\bibinfo  {journal} {JCAP}\ }\textbf {\bibinfo {volume} {05}},\
  \bibinfo {pages} {001}},\ \Eprint {https://arxiv.org/abs/0710.1302}
  {arXiv:0710.1302 [hep-th]} \BibitemShut {NoStop}%
\bibitem [{\citenamefont {Flauger}\ \emph {et~al.}(2013)\citenamefont
  {Flauger}, \citenamefont {Green},\ and\ \citenamefont
  {Porto}}]{Flauger:2013hra}%
  \BibitemOpen
  \bibfield  {author} {\bibinfo {author} {\bibfnamefont {R.}~\bibnamefont
  {Flauger}}, \bibinfo {author} {\bibfnamefont {D.}~\bibnamefont {Green}},\
  and\ \bibinfo {author} {\bibfnamefont {R.~A.}\ \bibnamefont {Porto}},\ }\href
  {https://doi.org/10.1088/1475-7516/2013/08/032} {\bibfield  {journal}
  {\bibinfo  {journal} {JCAP}\ }\textbf {\bibinfo {volume} {08}},\ \bibinfo
  {pages} {032}},\ \Eprint {https://arxiv.org/abs/1303.1430} {arXiv:1303.1430
  [hep-th]} \BibitemShut {NoStop}%
\bibitem [{\citenamefont {Shukla}\ \emph {et~al.}(2016)\citenamefont {Shukla},
  \citenamefont {Trivedi},\ and\ \citenamefont {Vishal}}]{Shukla:2016bnu}%
  \BibitemOpen
  \bibfield  {author} {\bibinfo {author} {\bibfnamefont {A.}~\bibnamefont
  {Shukla}}, \bibinfo {author} {\bibfnamefont {S.~P.}\ \bibnamefont
  {Trivedi}},\ and\ \bibinfo {author} {\bibfnamefont {V.}~\bibnamefont
  {Vishal}},\ }\href {https://doi.org/10.1007/JHEP12(2016)102} {\bibfield
  {journal} {\bibinfo  {journal} {JHEP}\ }\textbf {\bibinfo {volume} {12}},\
  \bibinfo {pages} {102}},\ \Eprint {https://arxiv.org/abs/1607.08636}
  {arXiv:1607.08636 [hep-th]} \BibitemShut {NoStop}%
\bibitem [{\citenamefont {Musso}(2013)}]{Musso:2006pt}%
  \BibitemOpen
  \bibfield  {author} {\bibinfo {author} {\bibfnamefont {M.}~\bibnamefont
  {Musso}},\ }\href {https://doi.org/10.1007/JHEP11(2013)184} {\bibfield
  {journal} {\bibinfo  {journal} {JHEP}\ }\textbf {\bibinfo {volume} {11}},\
  \bibinfo {pages} {184}},\ \Eprint {https://arxiv.org/abs/hep-th/0611258}
  {arXiv:hep-th/0611258} \BibitemShut {NoStop}%
\bibitem [{\citenamefont {Adshead}\ \emph {et~al.}(2009)\citenamefont
  {Adshead}, \citenamefont {Easther},\ and\ \citenamefont
  {Lim}}]{Adshead:2009cb}%
  \BibitemOpen
  \bibfield  {author} {\bibinfo {author} {\bibfnamefont {P.}~\bibnamefont
  {Adshead}}, \bibinfo {author} {\bibfnamefont {R.}~\bibnamefont {Easther}},\
  and\ \bibinfo {author} {\bibfnamefont {E.~A.}\ \bibnamefont {Lim}},\ }\href
  {https://doi.org/10.1103/PhysRevD.80.083521} {\bibfield  {journal} {\bibinfo
  {journal} {Phys. Rev. D}\ }\textbf {\bibinfo {volume} {80}},\ \bibinfo
  {pages} {083521} (\bibinfo {year} {2009})},\ \Eprint
  {https://arxiv.org/abs/0904.4207} {arXiv:0904.4207 [hep-th]} \BibitemShut
  {NoStop}%
\end{thebibliography}%

\end{document}